\documentclass[12pt,preprint]{aastex}

\begin{document}

\shortauthors{Mart\'inez-Sykora, et al.\ }

\title{Forward modeling of emission in SDO/AIA passbands from dynamic 3D simulations}

\author{Juan Mart\'inez-Sykora $^{1,2}$}
\email{j.m.sykora@astro.uio.no}
\author{Bart De Pontieu$^{1}$}
\author{Paola Testa$^{3}$} 
\author{Viggo Hansteen$^{2}$}

\affil{$^1$ Lockheed Martin Solar and Astrophysics Laboratory, Palo Alto, CA 94304}
\affil{$^2$ Institute of Theoretical Astrophysics, University of Oslo, P.O. Box 1029 Blindern, N-0315 Oslo, Norway}
\affil{$^3$ Harvard-Smithsonian Center for Astrophysics, Cambridge, MA 02138}

\newcommand{\myemail}{juanms@astro.uio.no}

\begin{abstract}  
It is typically assumed that emission in the passbands of
the Atmospheric Imaging Assembly (AIA) 
onboard the Solar Dynamics Observatory (SDO) is dominated by single
or several strong lines from ions that under equilibrium conditions
are formed in a narrow range of temperatures.
However, most SDO/AIA channels also contain contributions from lines
of ions that have formation temperatures that are significantly
different from the ``dominant" ion(s). We investigate the importance
of these lines by forward modeling the emission in the SDO/AIA
channels with 3D radiative MHD simulations of a model that spans 
the upper layer of the convection zone to the
low corona. The model is highly dynamic. In addition, we pump a steadily
increasing magnetic flux into the corona, in order to increase the
coronal temperature through the dissipation of magnetic stresses. As a
consequence, the model covers different ranges of coronal temperatures
as time progresses. The model covers coronal temperatures that are
representative of plasma conditions in coronal holes and quiet sun. 
The 131, 171, and 304~\AA{} AIA passbands 
are found to be least influenced by the so-called ``non-dominant"
ions, and the emission observed in these channels comes
mostly from plasma at temperatures near the formation temperature of the 
dominant ion(s). On the other hand, the other channels are strongly
influenced by the non-dominant ions, and therefore significant emission in 
these channels comes from plasma at temperatures that are
different from the ``canonical'' values. We have also studied the 
influence of non-dominant ions on the AIA passbands when
different element abundances are assumed (photospheric and coronal), 
and when the effects of the 
electron density on the contribution function are taken into account.
\end{abstract}


\section{Introduction}

The Solar Dynamics Observatory (SDO) was launched in February
2010.  The normal-incidence extreme-ultraviolet (EUV) and
ultraviolet (UV) telescopes of the Atmospheric Imaging Assembly (AIA)
onboard SDO provide full-disk
coverage in 9 different relatively narrow passbands every 12
seconds. One of the goals of AIA is to provide full thermal coverage
 in the photosphere, transition region and, especially, the corona.

AIA images have been and will be used extensively to 
allow tracing of the flow of mass and energy in the solar atmosphere.
For instance, \citet{De-Pontieu:2011lr} compared chromospheric
H$\alpha$ data with AIA images of the transition region and corona in the
304, 171 and 211 \AA{} passbands, and found that chromospheric jets
\citep[so-called type II spicules or rapid blueshifted
events][]{De-Pontieu:2007cr,Rouppe-van-der-Voort:2009ul} 
are often associated with heating of significant 
amounts of plasma to coronal temperatures. 
Similarly, \citet{Berger:2011qy} used SDO/AIA 
images of prominences to conclude that the temperatures in the 
'bubbles' and plumes, seen as dark features in optical observations 
with Hinode, are of the order $10^5-10^6$~K. 
\citet{Schmelz:2011fj} used AIA observations to perform a multithermal 
analysis of different cool loops and describe them as isothermal or multi-stranded 
depending on which channels are observed. \citet{Krishna-Prasad:2011kx} 
observed that propagating disturbances driven by jets in polar coronal regions
show different propagation velocities for the various passbands, which could shed light on whether
these disturbances are caused by flows or waves.

Clearly an accurate interpretation of AIA images is crucial to make
significant advances in tracing the spatio-temporal evolution of the
coronal plasma, and thus develop a better understanding of such
diverse topics as coronal heating, the structure of loops, the nature
of propagating disturbances, the formation and impact of prominences, etc.
The nature of AIA's EUV channels can complicate the
interpretation of AIA's coronal data. This is because, while the passbands of AIA's EUV channels
are relatively narrow and tuned to emission from strong lines of
single ions, the resultant images are not spectrally pure. Most of the
passbands are several \AA{} wide and contain emission from spectral
lines emitted by ions that have different formation temperatures than
those of the dominant ion. This can provide a significant challenge 
when interpreting the EUV images to trace the flow of mass and energy 
in the corona.

One approach to investigate the impact of ``non-dominant'' lines on AIA's
EUV passbands has been taken by \citet{ODwyer:2010yq} who used 
differential emission measure (DEM) curves that are considered 
representative of the thermal distribution in different
solar features (coronal hole, quiet sun, active region, and flaring
plasma), and synthesized the spectra in the SDO/AIA passbands using
CHIANTI \citep{Dere:2009lr}. They used the emergent spectra to 
investigate the relative contribution
of different spectral lines to the observed AIA emission for these
model DEMs.  

However, the solar atmosphere is highly dynamic and the use of
``typical'' DEMs may not accurately reflect the solar conditions. In
order to obtain a better knowledge of the importance of the various
spectral lines on the SDO/AIA channels, we take a different approach
here. We use snapshots from realistic 3D MHD numerical simulations,
calculate the emergent intensity (using the CHIANTI atomic database) of all significant spectral lines
within the AIA passbands, and investigate the importance of
``non-dominant'' lines for various solar coronal conditions.

In Section~\ref{sec:model}, we describe the Bifrost code used for
simulating the solar corona and the setup of the simulation. 
Section~\ref{sec:synthetic} describes the approximations used in order 
to calculate the synthetic EUV emission.
The results are shown in Section~\ref{sec:resu}, where we first describe
an example of quiet Sun observations with SDO/AIA 
(see Section~\ref{sec:obs}). The large variety of atmospheric stratifications in the 3D MHD model
due to the evolution of the model is shown in Section~\ref{sec:strat}.
In Section~\ref{sec:diffion} we discuss the importance of each of the 
spectral lines contributing significantly to the emission
in the AIA passbands, and we look at the temperature dependence 
of a channel by analyzing the emission coming from different temperature ranges 
in the model.  In order to study the
validity of our results for the actual SDO/AIA observations, we have also 
studied the effects of degrading the synthetic observations to SDO/AIA resolution (see
Appendix~\ref{sec:reso}).
In addition, we calculated the impact on the AIA emission 
from assuming photospheric and coronal abundances 
(Appendix~\ref{sec:abund}). In Appendix~\ref{sec:elec} we compare the
emergent total intensity and contribution of non-dominant ions to the
AIA passbands when we assume that the contribution function ($G(T)$)
is only dependent on temperature, or when the electron density dependence is taken into account
($G(T,n_{\rm e})$). We draw our conclusions in Section~\ref{sec:concl}.

\section{Numerical model and setup}
\label{sec:model}

We investigate the importance of the various spectral lines on the
SDO/AIA channels by constructing synthetic observations based on
forward 3D MHD models using the Bifrost code \citep{Gudiksen:2011qy}.
The model spans the solar atmosphere from the upper layer of the
convection zone to the low corona some 14~Mm above the photosphere. 
The model is highly dynamic, especially as we artificially
increase the amplitude of the magnetic flux in the corona in order to 
ensure that the modeled coronal temperatures become fairly high
towards the end of the simulation.

The Bifrost code is used to solve the full MHD equations on a staggered 
mesh, with non-LTE and non-grey radiative transfer with scattering, a
realistic equation of state, and 
conduction along the field lines, as described in detail in 
\citet{Gudiksen:2011qy}. In essence, this model attempts to be as
 ``realistic'' as possible by including as much relevant physics as
 possible within the constraints of computational technology.

\begin{figure}
  \includegraphics[width=0.9\textwidth]{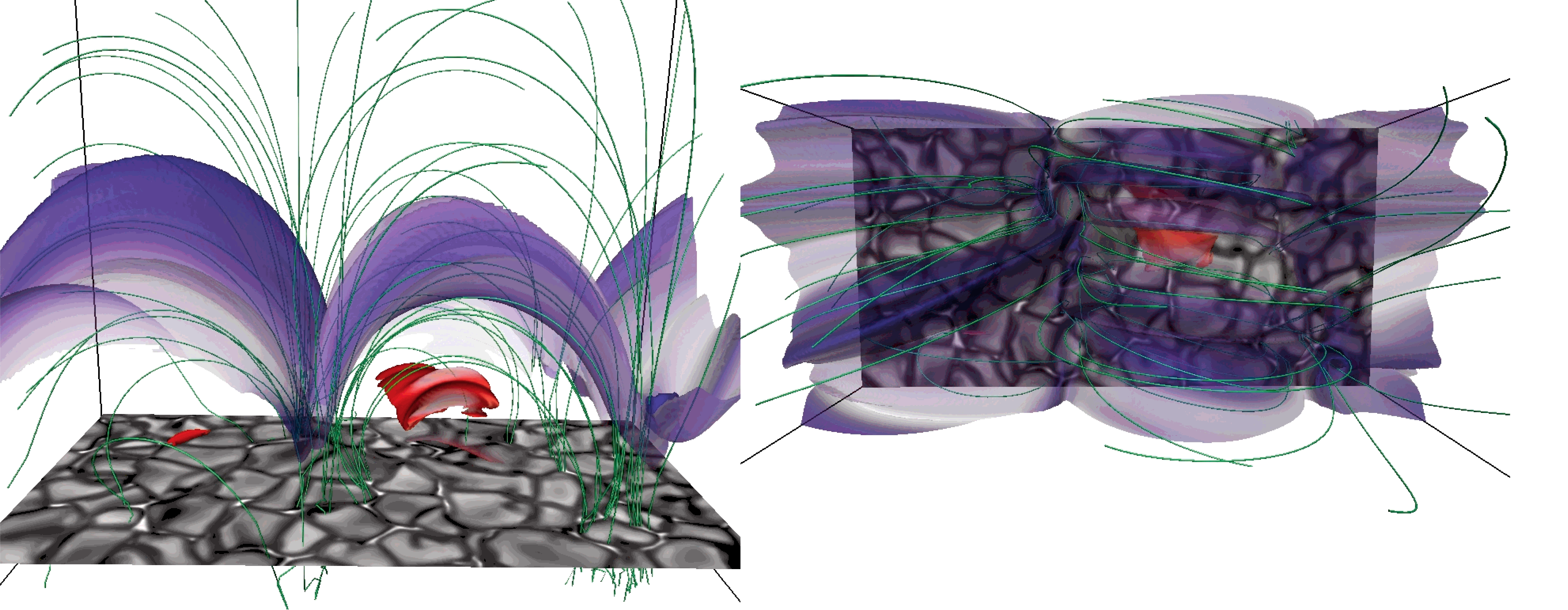}
 \caption{\label{fig:3dtgj2} A 3D snapshot of the 3D MHD model in a ``$xz$'' view 
 (left panel) and ``$xy$'' view (right panel) at time $t=1200$~s.
 The blue-red isosurface is at temperature $T=1.4\, 10^6$~K, 
 where blue for low density plasma
 ($\approx 8\,10^{-16}$ gr~cm$^{-3}$) and red for high density plasma ($\approx
 10^{-14}$ gr~cm$^{-3}$).  
 The photospheric continuum intensity is shown with the grey
 colortable, and some selected magnetic field lines are shown in green
 to give an indication of the magnetic field topology in the corona.}
\end{figure}

The computational domain stretches from the upper convection zone to the 
lower corona and is evaluated on a non-uniform grid of $512\times 256\times 360$
points spanning  $16\times 8\times 16$~Mm$^3$ implying a horizontal
grid size of $31$~km.  The frame of reference for the model
is chosen so that $x$ and $y$ are the horizontal directions (see Fig.~\ref{fig:3dtgj2}).
The grid is non-uniform in the vertical $z$-direction to ensure that
the vertical resolution is good enough to resolve the photosphere and
the transition region with a grid spacing of $28$~km, while becoming larger at 
coronal heights where gradients are smaller. 

The initial model is seeded with magnetic field, which rapidly receives 
sufficient stress from photospheric motions 
to maintain coronal temperatures in the upper part of the computational 
domain, in the same manner as first done by
\citet{Gudiksen+Nordlund2004}.
The model has an average unsigned field in the photosphere of
$160$~G. The magnetic field is distributed in the photosphere 
in two ``bands'' of vertical field centered around roughly $x=7$~Mm and
$x=13$~Mm.
In the corona this results in loop shaped structures, mostly oriented
in the $x$-direction, that stretch between these bands as shown in Fig.~\ref{fig:3dtgj2}.

We gradually increase the magnetic flux in the atmosphere in order to increase
the coronal temperature since the coronal heating in this model occurs
when stressed magnetic fields relax. The amplitude of the magnetic field 
is varied by the following expression:

\begin{eqnarray}
{\bf B_{new}}={\bf B_{ini}}(1+\delta/\tau)
\end{eqnarray}

\noindent where ${\bf B_{ini}}$ is the initial magnetic field in the box, and $\delta$ is 
the increment in the magnetic field flux, increasing every $\tau$ time
interval. Typically, $\delta=5\times 10^{-4}$ and $\tau=1$~s: this
time scale is large enough to let the model to relax and reach equilibrium
conditions before the next increase occurs. Thus, during the 1500~s of the
simulation, the magnetic field increases by a factor $0.75$.

\subsection{Synthetic data from SDO/AIA channels}
\label{sec:synthetic}

To analyze the emergent emission of the simulated atmosphere,
we calculate synthetic images for the various SDO/AIA channels.
The emission in each channel is calculated by computing the emission
in all lines lying in the corresponding narrow wavelength range of
channel sensitivity, folding the resulting spectrum with the wavelength
response of the channel and integrating in wavelength. This is done
for each vertical column in the numerical domain.
The emission for each line is calculated assuming the optically thin
approximation under ionization equilibrium conditions. Therefore, 
the intensity in a spectral line is calculated as follows: 

\begin{eqnarray}
I(S) = \int_{l} A_{b}\, n_{\rm e}(V)^{2}G(T(V),n_{\rm e}(V))dl. 
\end{eqnarray}

\noindent where $l$, $S$, and $V$ are length, area, and volume, 
respectively. In our calculations we obtain AIA synthetic images by
integrating ($l$) along the $z$ axis, therefore $S$ is the area in the 
$xy$ plane. $A_{b}$, $n_{\rm e}$, and $G(T,n_{\rm e}$)  
are the abundance of the emitting element, the electron density 
and the contribution function, respectively. 
The electron density is calculated using the Saha-Boltzman equation. 
We create a lookup table of the contribution function ($G(T)$ or 
$G(T,n_{\rm e})$) using the solarsoft package for IDL {\tt ch\_synthetic.pro}, 
where the keyword GOFT is selected. In general, we calculated the 
emission taking into account the density dependence of the transition, i.e., using a
contribution function $G(T,n_{\rm e})$ instead of $G(T)$. In cases where we use
$G(T)$ (see appendix~\ref{sec:elec}), a constant 
pressure is assumed ($P=10^{15} $cm$^{3}$K).
Knowing the temperature ($T$), and the electron density ($n_{\rm e}$) for each 
grid-point, $G(T,n_{\rm e}$) is obtained by interpolation in the lookup table. 
We do not calculate the chromospheric emission; the coolest ion that we take 
into account is \ion{He}{2} with a formation temperature $\log(T)=4.9$. To 
synthesize the plasma emission we use CHIANTI v.6.0.1 \citep{Dere:2009lr} 
with the ionization balance {\tt chianti.ioneq}, available in the CHIANTI
distribution. The characteristics of the SDO/AIA channels studied in the 
present work are listed in table~\ref{tab:chan}. We synthesized AIA 
observations for two different sets of  abundances: photospheric 
abundances \citep{Grevesse:1998uq}, and coronal abundances 
\citep{Feldman:1992qy}. 
We discuss in detail the effect of abundances in appendix~\ref{sec:abund}
but in the rest of the main text we will discuss results obtained
assuming photospheric abundances, unless noted otherwise.

\begin{figure}
  \includegraphics[width=0.5\textwidth]{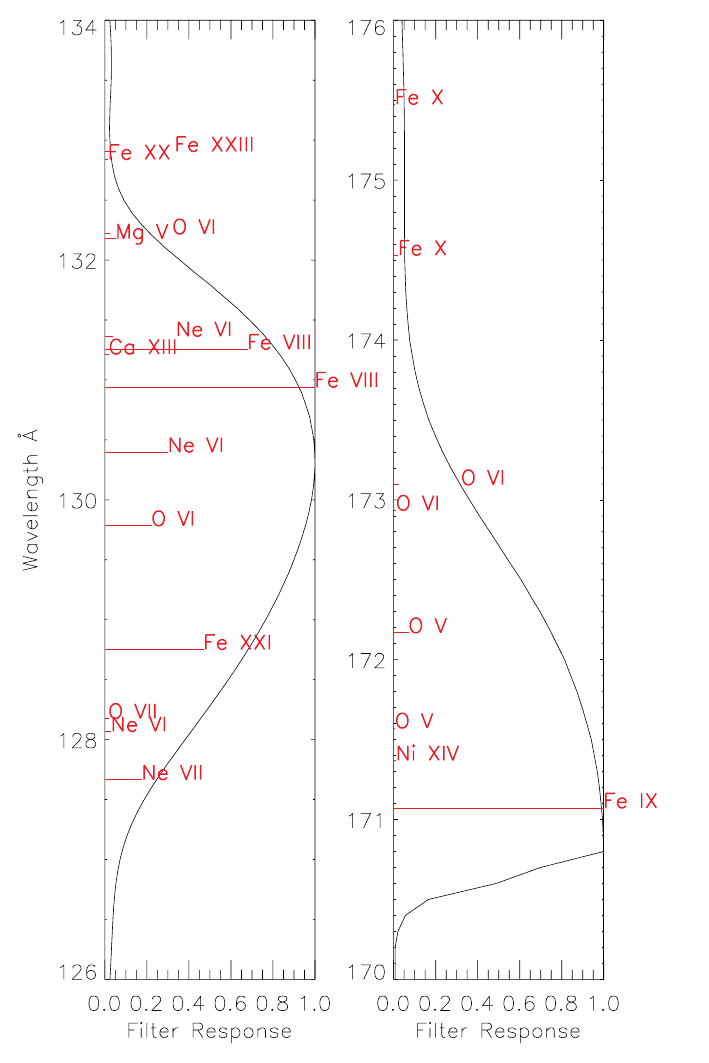}
 \caption{\label{fig:filtlines} Instrument response as a function of
   wavelength for the 131 (left panel)  and 171~\AA{} (right panel) channels. 
   These responses are available in the solarsoft package for IDL ({\tt aia\_get\_response.pro}). 
   The various spectral lines contribution within the passband are shown
   and labeled, and the length of the red line for each transition is proportional
   to the maximum of the contribution function for that transition (${\rm max}(G(T))$) (see 
   Section~\ref{sec:synthetic} for details of $G(T)$).}
\end{figure}

\begin{figure}
  \includegraphics[width=0.9\textwidth]{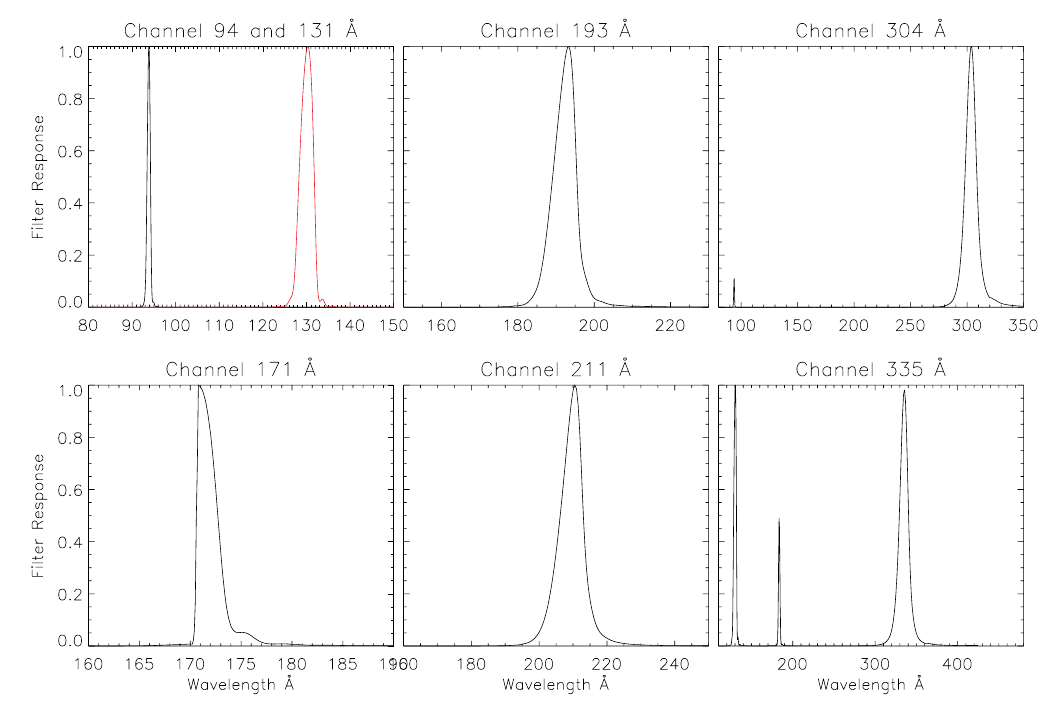}
 \caption{\label{fig:resp} Instrument response as a function of
   wavelength for the SDO/AIA channels analyzed in this work. 
   SDO/AIA EUV channels 94, 171, 193, 211, 304, and 335~\AA{} are
   shown from left to right and top to bottom respectively. The
   response of channel 131~\AA{} is shown in red in the top left
   panel together with channel 94~\AA{} (shown in black). 
   These response curves are available in the solarsoft package for
   IDL ({\tt aia\_get\_response.pro}).}
\end{figure}

As shown in Figure~\ref{fig:filtlines}, where we show the two filters
that are amongst the `cleanest', the passband of an AIA filter
contains many spectral lines from ions that are different from the dominant ion. 
We calculate the emission for all lines in the passband of each channel. 
The wavelength response of the EUV AIA channels are shown in
Figure~\ref{fig:resp}.  Note that the wavelength responses of the 304
and 335~\AA{} channels have secondary components, i.e.,  
non-negligible sensitivity in spectral ranges far from the central
wavelength ($\lambda$) of the passband. This is partly due to crosstalk with the
other channel on the same telescope, i.e., the 304~\AA{} channel 
is contaminated by the 94~\AA{} channel and the 335~\AA{} channel is
contaminated by the 131~\AA{} channel. 
The 335~\AA{} channel also has an additional component due to a
higher-order reflection from the multilayer mirror, around 185~\AA{}. 
Note that the width of the passband (in wavelength) is different for each channel,
ordered from narrower to wider: 94, 131, 171, 193, 211, 304, 335~\AA{}. 
Generally, contributions from a larger set of lines is more likely for
wider passbands.

\section{Results} \label{sec:resu}

\subsection{Observational example: quiet sun}  \label{sec:obs}

\begin{figure}
  \includegraphics[width=0.9\textwidth]{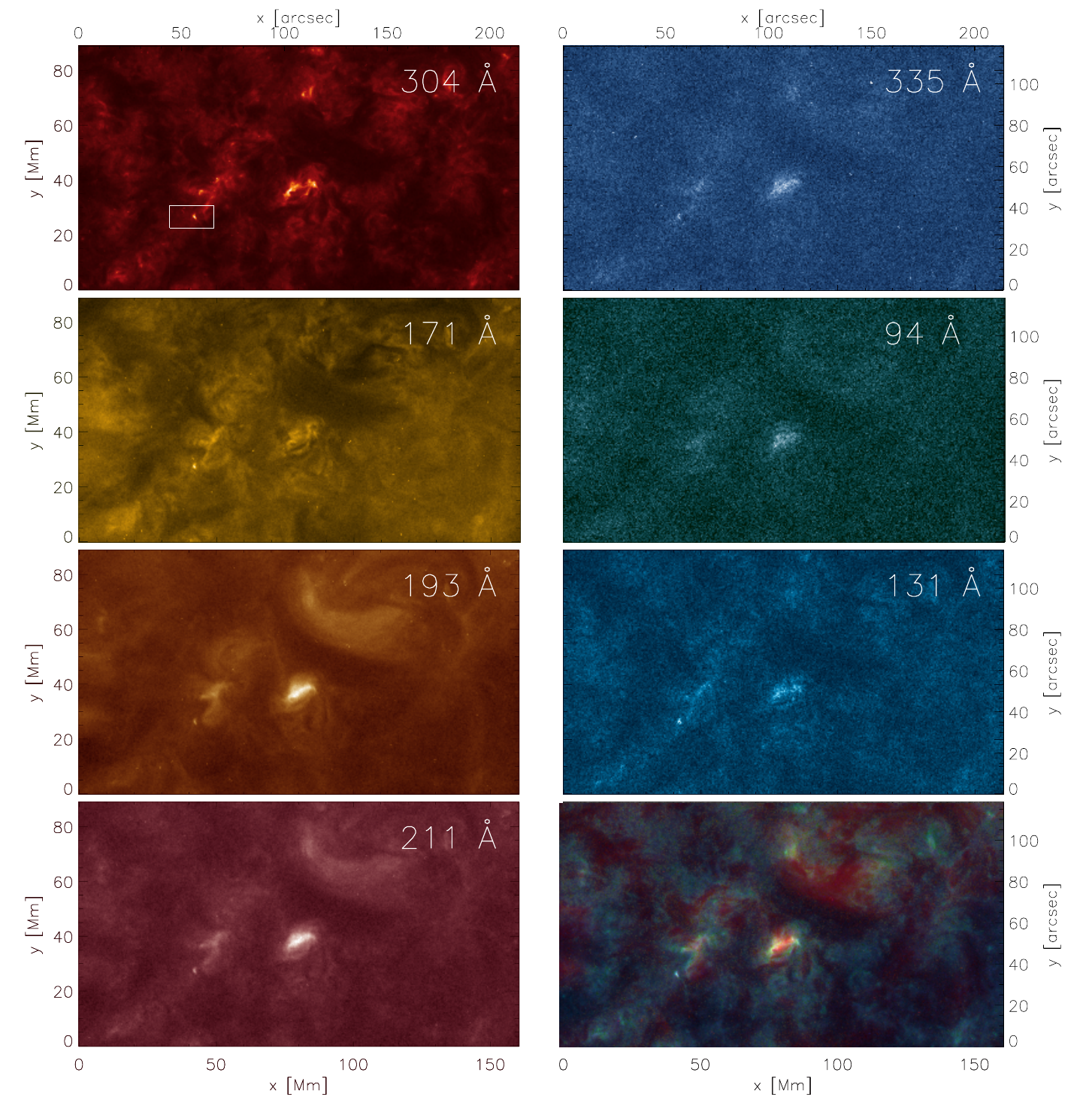}
 \caption{\label{fig:obs} Examples of SDO/AIA images of the 
 quiet sun taken at disk center on 2011 January 7 at 23:50 UT. 
 SDO/AIA channels 304, 171, 193, 211, 335, 94~\AA{}, 131~\AA{} 
 and a composite image with the 304~\AA{} (green), 193~\AA{} (red) 
 and 131~\AA{} (blue) channels are  shown, from top to bottom and left to right, respectively.
The white square in the top-left panel shows the size of the
simulated model.}
\end{figure}

SDO/AIA observes the full sun continuously with a cadence of $\sim
12$~s and at high spatial resolution ($0.6$~arcsec/pixel). 
Examples of recent AIA observations in the EUV channels that we simulate
are shown in Figure~\ref{fig:obs}. These data are observations of a
quiet sun region at disk center on 2011 January 7 around 23:50UT.

The passbands of EUV narrowband imaging instruments are often chosen
to contain a dominant single strong emission line or lines from the same
ion. Under these conditions, interpretation of the observations is much
simplified, as it can be assumed that most of the emission corresponds
to the dominant transition and is formed in a definite and narrow temperature range.  
However, in Figure~\ref{fig:obs}, one can note that some features are similar in 
most channels, suggesting that the same plasma is the source of the emission in all these
channels. An example of such a feature is the bright
point centered in the white box in the top-left panel.
Given the broad range of temperatures covered by the
dominant ions, this could mean that the plasma in the similar-looking
feature has a very broad thermal distribution. Alternatively, the
presence of lines from non-dominant ions in most of the AIA passbands
could, in principle, conspire to ``contaminate'' all passbands with
emission from one or a few ions that are formed at a narrow temperature range that
is different from what is expected from the canonical temperature
values for the passbands.
Here we investigate the importance of the contribution of
lines other than the dominant lines for a range of plasma
conditions. 

\subsection{Plasma stratification in the 3D model}
\label{sec:strat}

\begin{figure}
  \includegraphics[width=0.9\textwidth]{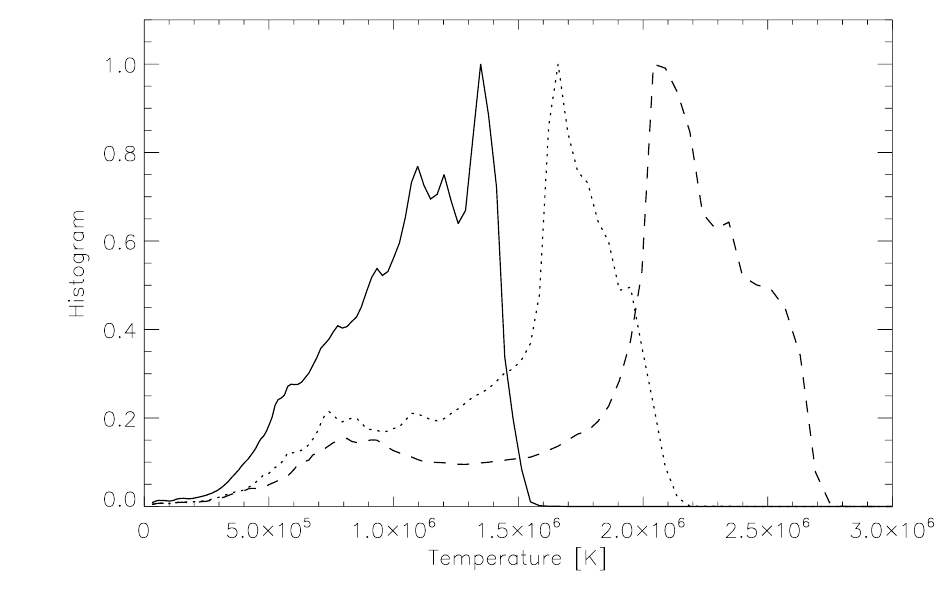}
 \caption{\label{fig:tempdist} Histogram of coronal temperatures of
   the model at t=[800, 1200, 1460]~s, shown with solid, dotted and
   dashed lines respectively. The histogram is for temperatures larger 
   than 30,000~K. The snapshots are representative of solar conditions
 in a coronal hole (800~s) and quiet Sun with small hotter emerging 
 regions (1200~s) and with hotter corona (1460~s).}
\end{figure}

As mentioned in section~\ref{sec:model}, we gradually increase the
magnetic flux in the model in order  to increase the magnetic field
stresses. This leads to a gradual increase with time of the average coronal
temperature.  
In order to simulate SDO/AIA observations in different average plasma
conditions, we select three different instants from the time-dependent
model, at $t=[800, 1200, 1460]$~s. The histogram of coronal temperatures at each of
the three times is shown in Figure~\ref{fig:tempdist}; at $t=800$~s 
the histogram peaks at $T=1.3$~MK, at $t=1200$~s around $1.7$~MK, and
at $t=1460$~s the peak is at a temperature above $2$~MK. This hottest
atmosphere has a maximum temperature of $\sim 2.8\, 10^6$~K. The
range of temperatures encountered in these different snapshots are
likely representative of solar conditions in a coronal hole
($t=800$~s) and quiet Sun, with small hotter emerging regions 
($t=1200$) and with a hotter corona (1460~s). 

\begin{figure}
  \includegraphics[width=0.9\textwidth]{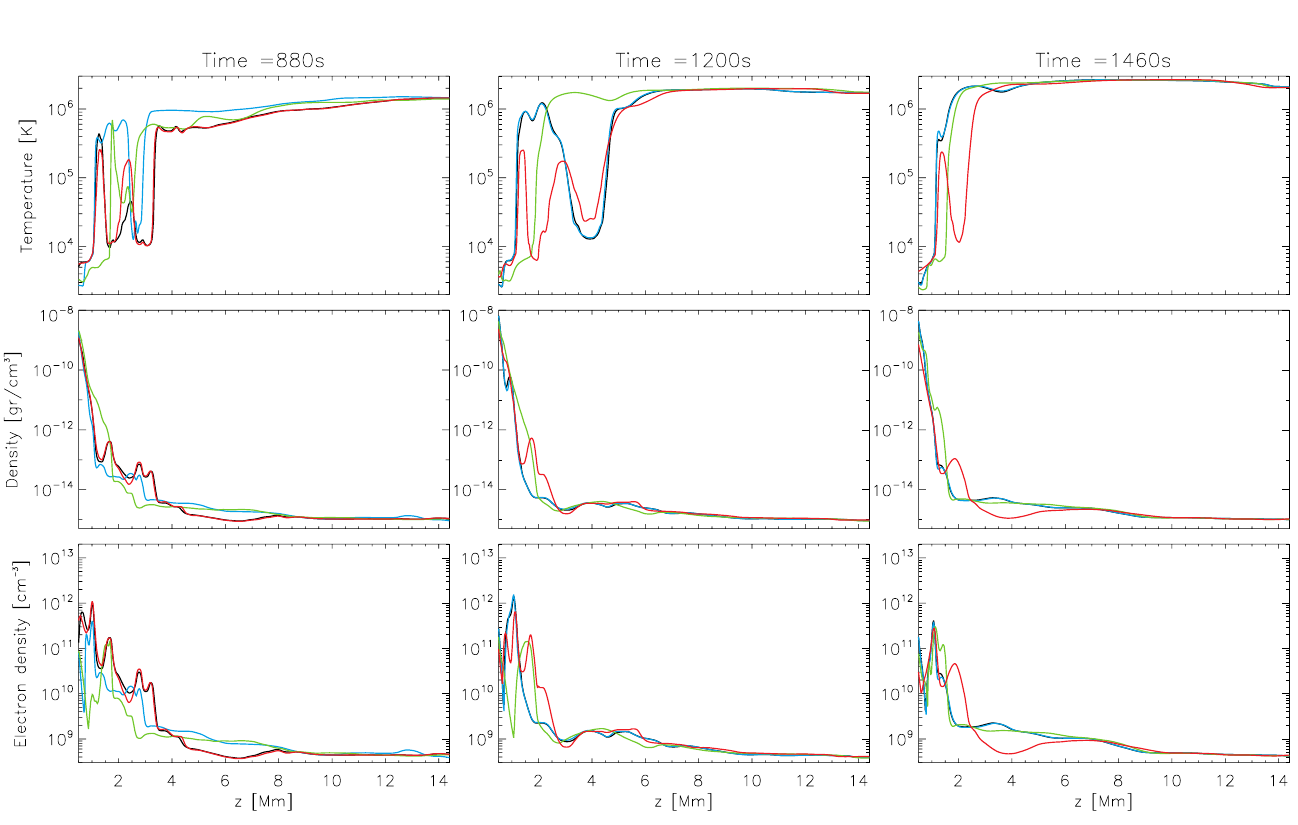}
 \caption{\label{fig:strat} Temperature, density and electron density as a function of 
 height are shown in each row from top to bottom, respectively. The columns from 
 left to right are at time 880, 1200 and  1460~s, respectively. 
 The red, blue, green and black lines correspond to the positions in $x$ and $y$
 that are marked with, respectively, dashed red, solid red, dashed green
 and solid green crosshairs in Figure~\ref{fig:aia211}.}
\end{figure}

Due to the various thermo-dynamics processes
and the natural evolution of the model, the temperature, density and electron density 
stratification vary considerably in both space and time, as shown in 
Figure~\ref{fig:strat} where selected `columns' of these variables are
plotted as a function of height $z$, for the three times used in 
this study.

The stratifications presented in Figure~\ref{fig:strat} show substantial differences.
For instance, some of the stratifications show a sharp transition region 
(green and black lines in the right panels). Other stratifications show cold 
and dense loops (red lines in the left panels). Moreover, a hot dense loop 
emerges into the corona around t=1200~s (blue and black lines in the 
middle column). One can appreciate this highly dense and hot loop
in the 3D images in Figure~\ref{fig:3dtgj2}. The physical processes and the
temporal evolution clearly differ in the different columns. Since these
processes are not the main interest in this work, we will not describe
them in detail here. However, discussions of similar simulations and the 
physical processes that govern them can be found in 
\citet{Hansteen+Carlsson+Gudiksen2007} and other papers. For instance, 
the propagating shocks which lead to type~{\sc i} spicules have been studied 
in detail by \citet{Hansteen+DePontieu2006,De-Pontieu:2007cr} and 
\citet{Martinez-Sykora:2009kl}. \citet{Hansteen:2010uq} observed from their realistic 
3D simulations a distribution of red-shifts with temperature that looks very similar 
to observations, and found that a heating mechanism that leads to episodic heating 
events in the vicinity of the upper chromosphere, transition region and lower corona 
naturally leads to a strong velocity gradient \citep{Martinez-Sykora:2011fj}. 
An important source of plasma injection into the corona are the jets and type~{\sc ii} spicules
\citep{Heggland:2009lr,Martinez-Sykora:2011uq}. In addition, the emergence of magnetic flux 
into the corona has been described by \citet{paper1,Martinez-Sykora:2009rw}. 

We will show in the following sections that as a consequence of these
and similar processes and their resulting stratifications, 
the emission in the SDO/AIA channels may often come from non-dominant
ions and/or from plasma that is significantly cooler or hotter than the formation
temperature of the dominant ion(s). 

\subsection{``Narrow'' filter passband description} \label{sec:diffion}

\begin{figure}
  \includegraphics[width=0.8\textwidth]{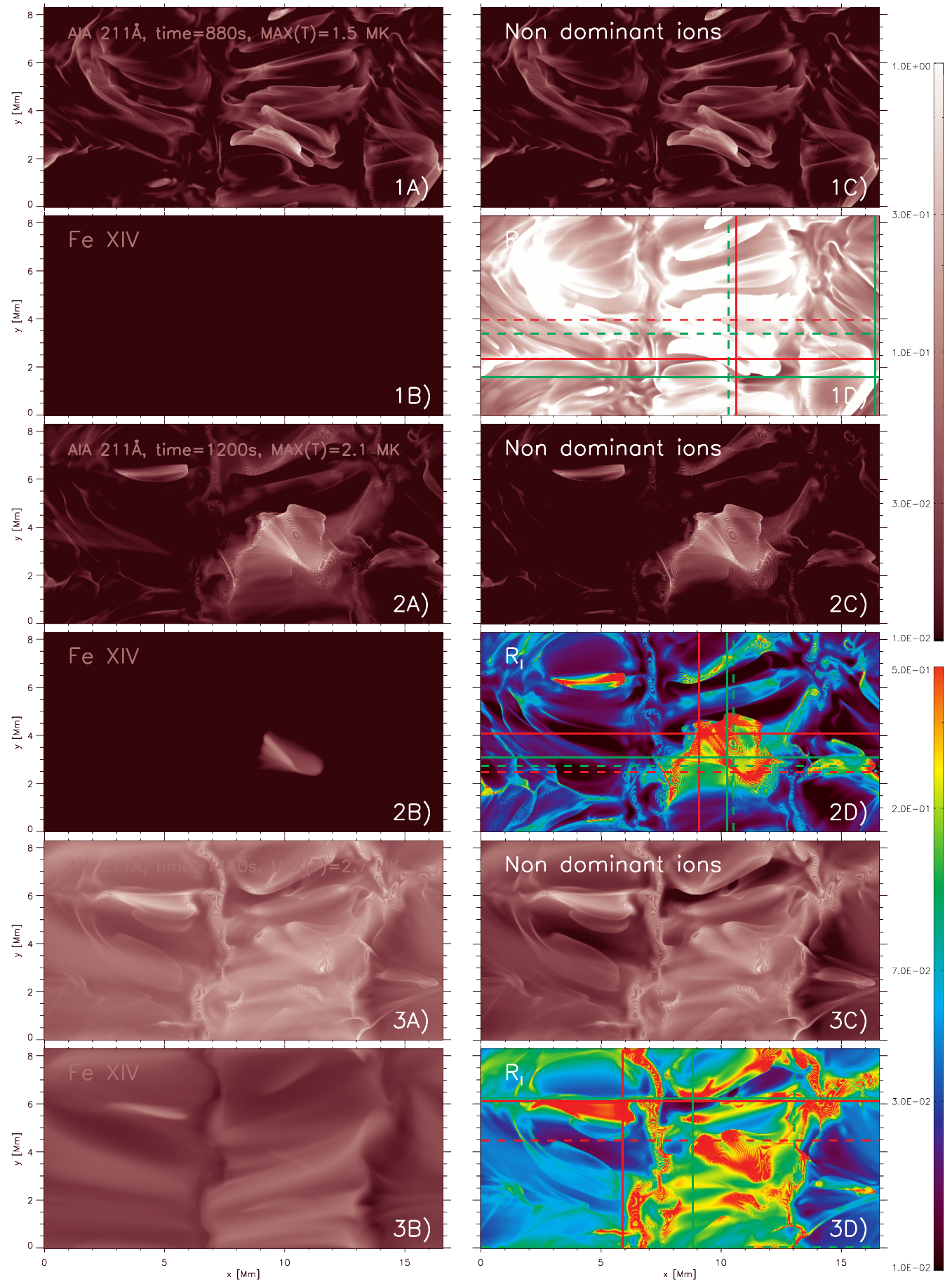}
 \caption{\label{fig:aia211} Synthetic images of channel 211~\AA{}. This 
collection of panels are set in three groups of four panels. Each
group is set a different time (880, 1200 and 1460~s) which are shown in the panels
labeled 1A-D, 2A-D and 3A-D, respectively. In each 
set of four panels, we show $I_T$, $I_D$, $I_{ND}$ and $R_I$ 
in panels labeled 1-3A,  1-3B, 1-3C, and 1-3D, respectively. The 
synthesized intensities have been normalized to the maximum intensity of the 
synthesized SDO/AIA channel (panels 1-3A) and all use the same color 
scheme in logarithmic scale (top colorbar). The color scheme for $R_I$ is shown 
with the bottom colorbar in logarithmic scale. 
The solid red, dashed red, solid green and dashed green crosshairs in panels 1-3D
are the locations in $y$ corresponding to the panels labeled 1-3a, 1-3b, 1-3c 
and 1-3d in Figure~\ref{fig:contrib} and in Movie~2.}
\end{figure}

\begin{figure}
  \includegraphics[width=0.9\textwidth]{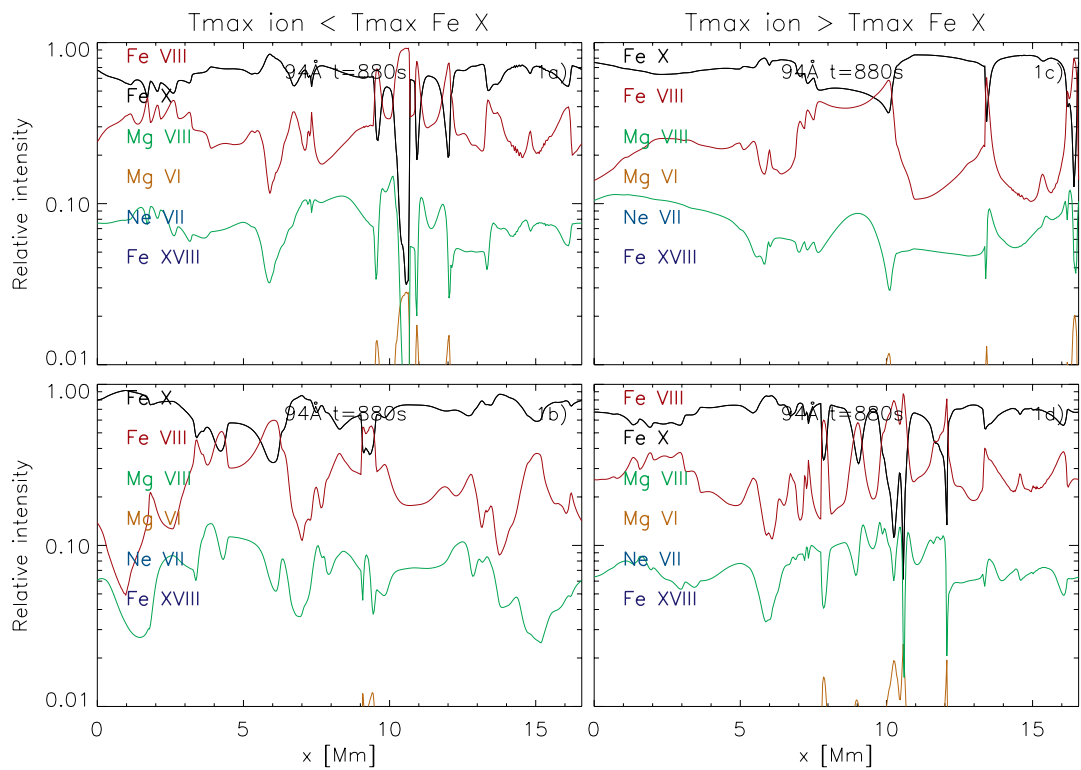}
 \caption{\label{fig:contrib}  Relative contribution 
for the most significant spectral lines in 
the 94~\AA{} channel, at time 880~s  
for four specific $y$-positions in the field of view. 
The ions listed in each plot are those with the most 
important contributions, ordered from the largest to the smallest contribution
(from top to bottom). Each panel shows a different $y$
position, which is marked with a crosshair in panel 1D of Figure~\ref{fig:aia211}. 
The solid and dashed red, and solid and dashed green crosshairs 
are shown in the panels labeled 1a, 1b, 1c, and 1d in this figure,
respectively.}
\end{figure}

\begin{figure}
  \includegraphics[width=0.75\textwidth]{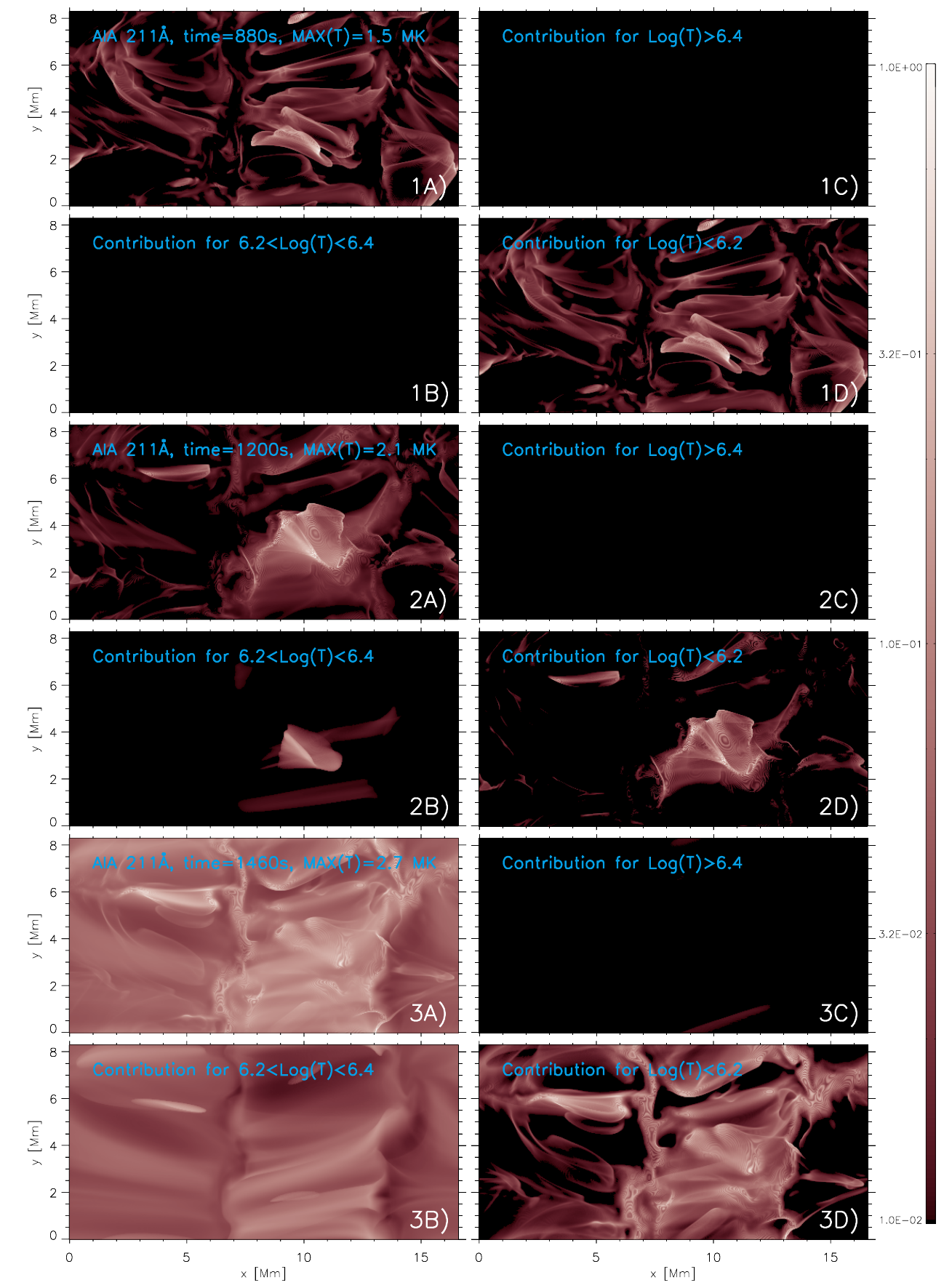}
   \caption{\label{fig:temp} Emission contribution for different temperature ranges
for the synthesized 211~\AA{} emission is shown at $t=880$~s (panels
labeled 1A-D), $t=1200$~s (panels labeled 2A-D), and at $t=1460$~s
(panels labeled 3A-D).  The synthesized intensities $I_T$, $I_I$,
$I_H$, and $I_L$ are shown in the panels labeled 1-3A, 1-3B, 1-3C and
1-3D, respectively (see text for definition). 
The color-scheme is shown in the right 
colorbar in logarithmic scale and all panels are normalized to the
maximum of the synthesized intensity (panels 1-3A).}
\end{figure}

While we discuss all SDO/AIA channels in this section, we
only show figures for one channel in the main paper (211~\AA\, see 
Figure~\ref{fig:aia211}-\ref{fig:temp}). However, 
similar figures for all channels can be found in the on-line version 
of the paper (Movies~1-3). Figure~\ref{fig:aia211}
and the equivalent figures in Movie~1 
are composed of three groups of four panels. Each group corresponds
to a different snapshot in time (labeled as 1A-D, 2A-D and 3A-D at times 880, 1200 and
1460~s, respectively).  As mentioned, these three different snapshots
have coronae with very different temperature distributions; in general
the coronal temperature increases with time as shown in Figure~\ref{fig:tempdist}.
The four panels of each group (or each image in 
Movie~1) show: the total synthesized emission ($I_T$, labeled with
``A''), the intensity emitted by the dominant ion(s) ($I_D$, labeled
with ``B''), the intensity emitted in spectral lines of non-dominant ions
($I_{ND}$, labeled ``C''), and the ratio $R_I$ (labeled with ``D'').
For each SDO/AIA channel, $R_I$ is the ratio between the emission from the non-dominant
ions and the emission from the dominant ion(s), weighted with the synthesized intensities
for the specific channel, as defined in the following expression:

\begin{eqnarray}
R_I=\frac{I_{ND}}{I_D}I_T
\end{eqnarray}

The color scale for the panels 1-3D is shown at the bottom-right of
Figures~\ref{fig:aia211}, or at the top-right in the movies. 
Orange-red colored areas correspond to regions where the emission from
the non-dominant ions is relatively large compared to the emission from
the dominant ions, and in addition where the total emission of the synthesized 
channel is significant. The color-scale for panels 1-3A, B and C is shown in 
the top-right of Figure~\ref{fig:aia211} or the top-left of Movie~1. The 
color-scheme of this color bar changes from channel to channel, to
allow easier identification of the channels in Movies~1, 3, 4, 5, and
6. We use the same color schemes as those of the SDO/AIA webpages 
({\tt http://sdo.gsfc.nasa.gov/data/}). These color schemes can be loaded
using the solarsoft routine {\tt aia\_lct.pro}.

\begin{figure}
  \includegraphics[width=0.9\textwidth]{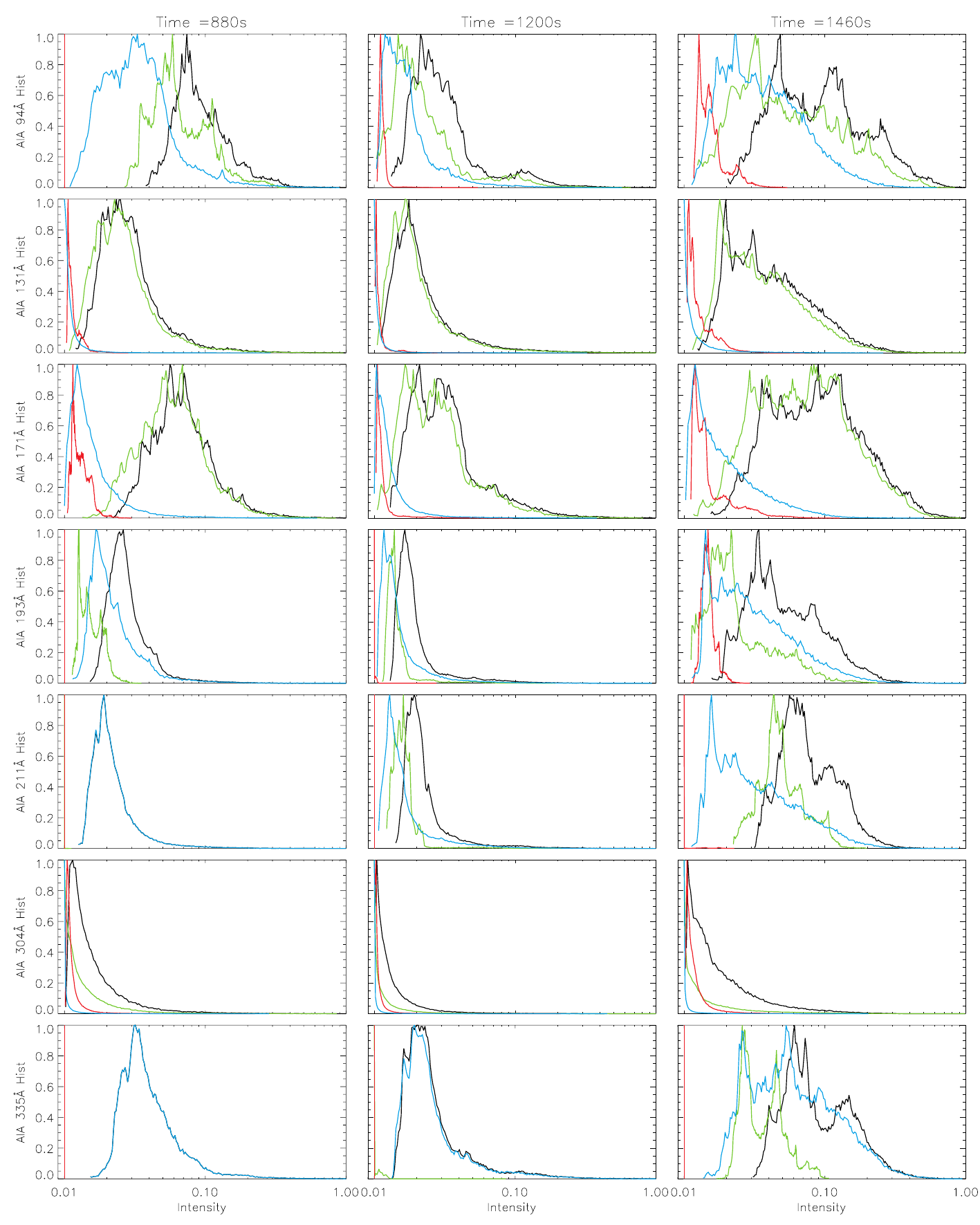}
 \caption{\label{fig:inthis}  Histograms of the intensities $I_L$,  
 $I_H$, $I_I$, and $I_T$ are shown in 
 blue, red, green and black lines. 
The columns are at the timesteps 
 880~s, 1200~s and 1460~s 
 from left to right, and the rows are the different SDO/AIA channels.}
\end{figure}

\begin{figure}
  \includegraphics[width=0.9\textwidth]{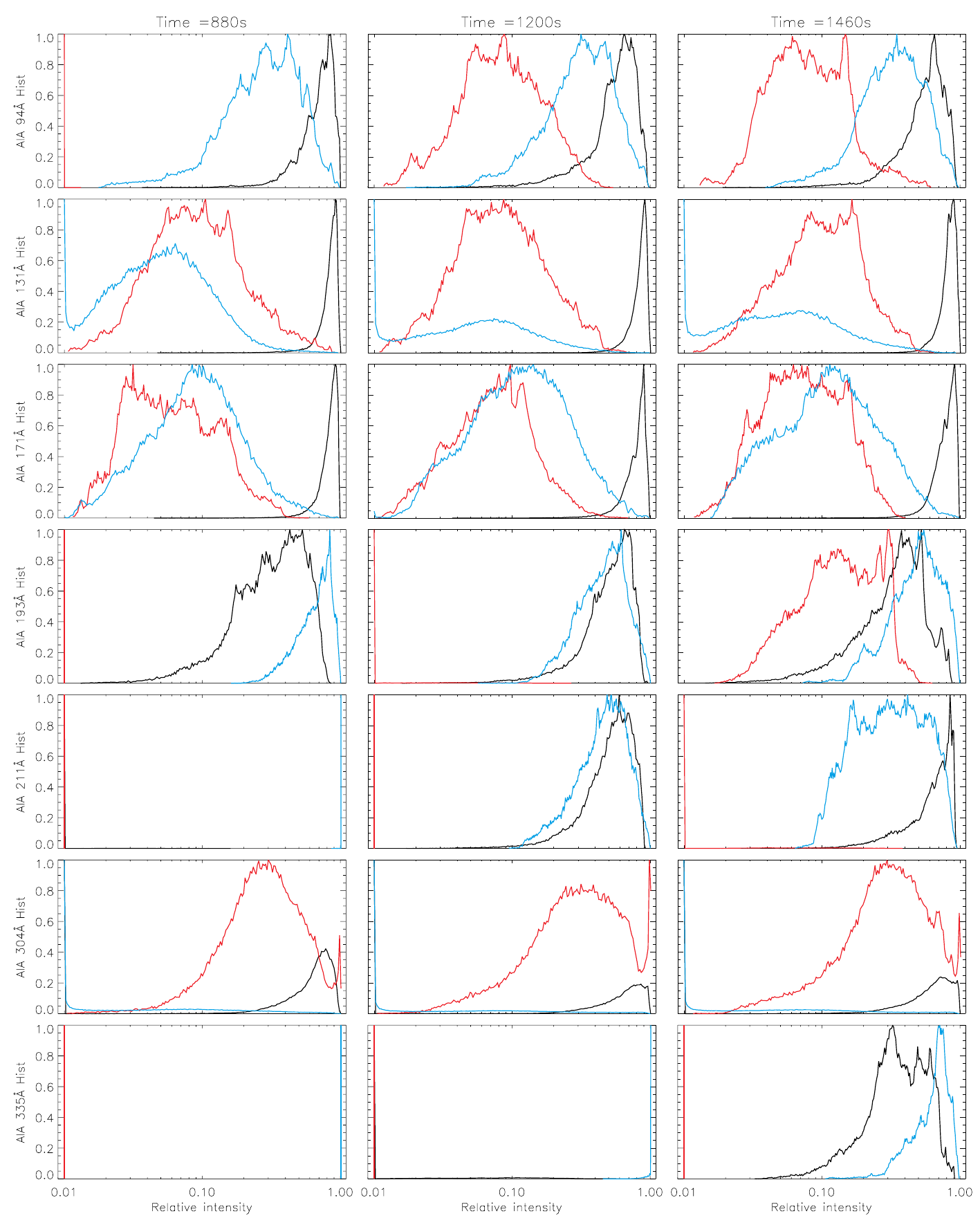}
 \caption{\label{fig:conthis}  Histograms of the relative contribution from 
$I_L$/$I_T$,  $I_H$/$I_T$, and $I_I/I_T$ are shown in blue, 
red and black lines. The columns 
are at the timesteps 880~s, 1200~s and 1460~s from left to right, 
and the rows are the different SDO/AIA channels.}
\end{figure}

Before discussing in detail the plots shown in Figures~\ref{fig:aia211}, 
and Movie~1 for each channel, we describe Figures~\ref{fig:contrib} 
and~\ref{fig:temp} and the corresponding Movies~2 and 3, and 
Figures~\ref{fig:inthis} and~\ref{fig:conthis}.  In Figure~\ref{fig:contrib}, 
and in each image in Movie~2, we plot the relative contribution of the 
most significant spectral lines from the various ions within a passband.
This shows how much emission comes from each ion (normalized to the
total emission) in a given filter for a specific $y$-position. In each plot we list the
ions with the most significant contributions, ordered from the one with 
the largest relative contribution along the x-axis (top) to the one
with the lowest relative contribution (bottom). In order to clarify: 
if the emission only came from the dominant ion, its relative 
contribution would be equal to one everywhere. We show results for four 
different $y$-positions within the field of view (one per panel). The 
different positions in $y$ are marked with differently colored and 
styled crosshairs in the panels 1-3D in Figure~\ref{fig:aia211} 
and in Movie~1. The relative contribution for the locations marked by 
the solid red, dashed red, solid green and dashed green crosshairs 
are shown in the panels labeled 1-3a, 1-3b, 1-3c and 1-3d in Figure~\ref{fig:contrib} 
and in Movie~2. We define $I_{NDL}$ as the intensity coming from the non-dominant
ions that have a peak formation temperature lower than that of the dominant
ion, and $I_{NDH}$ as the intensity coming from the non-dominant ions 
which have a peak formation temperature higher 
than that of the dominant ion. With these definitions, the location of each crosshair 
is then selected as follows:
the solid red crosshair is where $(I_{NDL}/I_D)I_T$ is largest, i.e., 
most of the emission comes from ions that have formation temperatures lower than the
dominant ion (at locations where the emission is not negligible); 
dashed red crosshair is where $I_{NDL}/I_D$ is largest, i.e., 
most of the emission comes from ions that have formation temperatures lower than the
dominant ion;  
solid green crosshair is where $(I_{NDH}/I_D)I_T$ is largest, i.e.,
most of the emission comes from ions that have formation temperatures higher than the
dominant ion (at locations where the emission is not negligible); 
and dashed green crosshair is where $I_{NDH}/I_D$ is largest, i.e.,  
most of the emission comes from ions that have formation temperatures higher than the
dominant ion (see Table~\ref{tab:cross}). 

Using the same synthetic calculations, we can look at the same data,
but categorize the emission for each channel as a function of temperature of 
the emitting plasma, instead of in terms of the ``dominant" and ``non-dominant" 
ions. Figure~\ref{fig:temp}  shows the emission coming from various 
temperature ranges in the model for the 211~\AA{} channel at $t=880$~s 
(top four panels labeled 1A-D), at $t=1200$~s (middle four panels 
labeled 2A-D), and at $t=1460$~s (bottom four panels 
labeled 3A-D). Each group of four panels and each image
in Movie~3, where we present similar plots for the other channels, 
has the same layout. The synthesized image for the 
specific channel is shown in the panels labeled 1-3A. 
For each passband three temperature
ranges are defined. The emission is then calculated, taking into account all the various spectral 
lines within the passband. The intensity coming 
from plasma at temperatures where the $G(T,n_{\rm e})$ of the dominant ion is
larger than
$1/e$ of its maximum is shown in the panels 1-3B ($I_I$). For example,
for the 211~\AA{} channel the corresponding temperature range is $6.2<\log(T)<6.4$
(see Figure~\ref{fig:temp}). For the channels with more than one dominant ion, 
we choose the dominant ion which has the lowest formation temperature, 
since the temperatures of our model are in the lower range of coronal temperatures 
observed on the sun. 
Panels 1-3C show the intensity $I_H$ coming from plasma at temperatures higher than the
high temperature value where the $G(T,n_{\rm e})$ of the dominant ion is $1/e$ of
its maximum, e.g., for the 211~\AA\ channel $I_H$ is the intensity of emission
from plasma at temperatures higher than $\log(T)=6.4$. Finally, panels 1-3D
show the intensity $I_L$ coming from plasma at temperatures lower than the low
temperature value where the $G(T,n_{\rm e})$ of the dominant ion is 
$1/e$ of its maximum, e.g., from plasma at temperatures lower than $\log(T)=6.2$ 
for the 211~\AA\ channel.

In order to quantify the dependence of the emission with temperature we show 
Figures~\ref{fig:inthis} and~\ref{fig:conthis}. Histograms of the
intensities and the relative intensities coming 
from different temperature ranges are shown in Figure~\ref{fig:inthis} and 
Figure~\ref{fig:conthis}, respectively. The histograms of the intensities $I_L$, $I_H$, $I_I$, 
and $I_T$ are shown with blue, red, green and black lines in Figure~\ref{fig:inthis}. 
When the green line is close to the black line, most of the emission 
for a specific channel comes from temperatures near the formation temperature of the 
dominant ion. These histograms are shown for each channel (rows) and at the three 
different instants; $t=880, 1200, 1460$~s (columns). 

The histograms of the relative intensities $I_L$/$I_T$,  $I_H$/$I_T$, and 
$I_I/I_T$ are shown with blue, red and black lines
in Figure~\ref{fig:conthis}. As in Figure~\ref{fig:inthis}, the columns 
are for the timesteps $t=880, 1200,1460$~s from left to right, with
the different SDO/AIA channels in different rows. Note that when the black
histogram is concentrated on a relative intensity of 1, all 
the emission comes from temperatures near  
the formation temperature of the dominant ion. When the black
histogram is more spread out, other temperature ranges also contribute
significantly to the passband. 

As shown in detail in the appendix~\ref{sec:reso}, we also investigated the 
effect of the instrumental spatial resolution by degrading the synthetic 
emission to the AIA spatial resolution. In summary, we found that the 
detailed description given for each channel 
below is also valid when the synthesized intensities are convolved and 
pixelized to the SDO/AIA resolution. However, the smaller structures disappear 
or are mixed with emission coming from other structures.  

\subsubsection{94~\AA{} channel}

The SDO/AIA 94~\AA{} channel has emission from two dominant ions with rather
different formation temperatures; \ion{Fe}{10} at $10^6$~K and
\ion{Fe}{18} at $6\times 10^6$~K (see Table~\ref{tab:chan}). 
Moreover, there is some evidence suggesting that this passband
contains emission from some lines that are not included in CHIANTI
\citep[e.g.,][]{Testa:2011fk,Foster:2011fk,Schmelz:2011fj,Aschwanden:2011fj},
therefore our analysis of the contributions to the 94~\AA\ emission
might be inaccurate. However, until the impact of these missing transitions 
will be assessed and their emission properly modeled and included in the 
atomic databases, the interpretation of AIA 94~\AA\ observations will be 
based on existing atomic data, and therefore we believe that our investigations
using the CHIANTI database are relevant for present AIA studies.
Our model does not reach extremely high temperatures, so
very little emission comes from the \ion{Fe}{18} lines. As a result, we
consider \ion{Fe}{10} as a dominant ion. At any given time 
non-dominant ions contribute significantly changing both the total 
intensity and shape of observed features (see panels 1-3C and 1-3D in Movie~1
for the corresponding channel). 
For example, at time $t=1200$~s (panels 2A-D), a bright emerging loop appears
roughly in the center of the image. This structure shows a significantly 
different intensity in  the synthesized 94~\AA{} channel (panel 2A in 
Movie~1 for the corresponding channel) as compared to the intensity
from the dominant ions (see panel 2B). This is because the loop has a 
strong contribution from ions with formation temperatures cooler than 
that of the dominant ions (see left panels in the corresponding image
in Movie~2).  Note that while the intensity is different, the shape of this 
emerging hot dense loop is similar in both 
images (panels 2A and 2B in Movie~1).

The non-dominant ions that provide the most important contributions to
the 94~\AA\ emission are different from gridpoint to gridpoint and from timestep to 
timestep due to the highly dynamic nature of the atmosphere. This can 
be seen in the corresponding snapshots of Movie~2. It is important to 
remark that the relative contribution of each ion changes considerably for
different positions in space in the same snapshot (i.e., at the same instant).
For instance, panels 1-3a and 1-3b in Movie~2 include the position where 
non-dominant ions that have formation temperatures lower than the 
dominant ion provide strong contributions to the emission. In these positions, 
the most important ion is \ion{Fe}{8}. Emission from other ions is at least one 
order of magnitude smaller than the emission coming from the dominant 
ion with, in order of importance, \ion{Mg}{8}, \ion{Mg}{6}, and \ion{Ne}{7}. 
The dominant ion \ion{Fe}{10} in these positions is also one order of 
magnitude smaller than the most important ion (\ion{Fe}{8}). Panels 
1-3c and 1-3d in Movie~2 include the locations where non-dominant ions with 
formation temperatures higher than the dominant ion provide strong 
contributions, i.e., $(I_{NDH}/I_D)I_T$ and  $I_{NDH}/I_D$  are highest. 
In these positions, the most important ions are \ion{Fe}{8}, 
\ion{Fe}{10}, \ion{Mg}{8}, \ion{Mg}{6}, and \ion{Ne}{7} in order of
importance, at the first timesteps (1a-d and 2a-d in Movie~2). Note that \ion{Fe}{8}, 
\ion{Mg}{8}, \ion{Mg}{6}, and \ion{Ne}{7}  have a formation temperature lower 
than of the dominant \ion{Fe}{10} ion, i.e.,  in locations where hotter ions play 
the most important role they are still insignificant compared to the ions with a 
lower temperature formation taking into account the temperature range of the 
simulation. 

It is interesting to note that the relative contribution from the various ions 
varies by orders of magnitude in space (as shown, e.g., in the horizontal direction
along $x$ in Movie~2).  Therefore, when we observe the Sun,
the most important relative contribution to this channel may be
from different ions depending on exactly when and where we are looking,
and can be very different at locations only a few pixels apart.
When the model gets hotter, a different set of ions become more important
(3a-d, Movie~2), such as \ion{Fe}{18}. However, the impact of the
non-dominant ions is generally less important when the model reaches higher 
temperatures (compare the panels 3A-D with the panels 2A-D and 1A-D 
in Movie~1). In sumary, the most important ion in the channel varies between 
\ion{Fe}{8}, \ion{Fe}{10}, and \ion{Fe}{18}, with the other ions having 
contributions that are for the most regions one order of magnitude smaller.

In panels 1-3B of Movie~3 we show the emission in the 94~\AA{} channel
from the plasma with a range in temperature of $\log(T)=[5.9,6.2]$. 
At $t=880$~s, an important contribution to the emission comes 
from plasma at temperatures cooler than $\log(T)=5.9$ (see 
panel 1C in Movie~3). This emission is
significant enough to impact the total 
synthesized intensity. As a result, some features that are 
observed in the emission from the range of temperatures 
$5.9<\log(T)<6.2$ have a different shape and intensity in the synthesized image of the
AIA channel (panels 1A in Movie~3). Comparing panels 2A-D and
3A-D with panels 1A-D it is clear that the influence of plasma cooler
than $\log(T)=5.9$ is less pronounced at later times. Most of the contributions
at temperatures cooler than $\log(T)=5.9$ are due to non-dominant ions 
which have formation temperatures lower than that of the
dominant ions \ion{Fe}{10} and \ion{Fe}{18}.

There is no emission coming from plasma at temperatures larger than $\log(T)=6.2$ 
at $t=880$~s (see panel 1C in Movie~3) since the model
does not reach temperatures greater than $\log(T)=6.2$ at this time. Later, when 
the model has reached temperatures that are larger than $\log(T)=6.2$,
such emission exists, but is generally negligible (see panels 2-3C in Movie~3 and top row 
in Figure~\ref{fig:inthis}). 

Figures~\ref{fig:inthis} and~\ref{fig:conthis} quantify the emission coming from 
the various temperature ranges. Most of the values for the normalized intensity 
$I_T$ are of order 0.08, 0.02, and 0.05 of the maximum intensity in the box at 
times 880~s, 1200~s, and 1460~s, respectively (see the black line
in the top row in Figure~\ref{fig:inthis}). At time 1200~s, the
emerging loop shows very strong emission which pushes the maximum
intensity in the box to very high levels. As a result, most of the
normalized intensities are quite small compared to its maximum, hence
the shift of the black histogram towards smaller values.

Ideally, when all the emission comes from within the temperature range 
$\log(T)=[5.9,6.2]$, the green lines coincide with the black lines. However,
in this channel, the black line is mostly a combination of the green
and blue line, i.e. the emission stems from plasma at the
temperature of the dominant ion and cooler.  
Most of the intensity coming from temperatures between
$5.9<\log(T)<6.2$ is at levels of 0.07, 
0.015, and 0.03 of the maximum intensity at time 
880~s, 1200~s and 1460~s, respectively
(see green lines in the first row in Figure~\ref{fig:inthis}). 
The intensity coming from temperatures lower than 
$\log(T)=5.9$ is found mostly to have values of 0.03, 
0.012 and 0.02 of the maximum at time 880~s, 
1200~s and 1460~s, respectively (see blue lines in the first row 
in Figure~\ref{fig:inthis}).  Contribution of cold plasma 
(blue lines) is roughly half of the contribution of the plasma 
at temperatures typical of the dominant ion (green lines). 
Therefore, the number of points emitting at intensities around the maximum 
in the histogram at time 880~s (black lines) is roughly 50\% 
larger than green line. For larger intensities (i.e., brighter regions),
at all instants, this difference between the green and black lines is
smaller. In the first row in Figure~\ref{fig:conthis} one can observe that in general 
$1/3$ or more of the emission comes from cold plasma (blue lines) and $2/3$ from the 
range of temperatures  $5.9<\log(T)<6.2$ (black lines). 

In appendix~\ref{sec:abund} we compare the emission using photospheric
and coronal abundances. This channel shows wide areas where the emission 
is different by assuming photospheric or coronal abundances, reflecting the
ratio of the abundance between the two datasets where Fe is different from Ne, 
which these two ions contribute to the emission. 
In appendix~\ref{sec:elec} 
we study, for each channel, the dependence of the emission on the
electron density. The 94\AA{} channel shows some significant dependence 
on the electron density because the \ion{Fe}{10} 94~\AA{} line is density 
sensitive and contributes significantly to the emission. 

In summary, the emission from cooler plasma plays a role everywhere
at all times in this channel, but it typically contributes less in brighter
and hotter regions.

\subsubsection{131~\AA{} channel}

The SDO/AIA 131~\AA{} channel has emission from two dominant ions which are formed at
completely different temperatures: \ion{Fe}{8} at $6\times 10^5$~K and
\ion{Fe}{20} at $10^7$~K; however, in non-flaring conditions emission
from the very 
hot \ion{Fe}{20} ion is negligible. The various structures observed 
in the synthetic 131~\AA{} channel  (panels 1-3A in Movie~1) are similar 
to those seen in the emission from the dominant ion (panels 1-3B in Movie~1).
Due to the temperatures that the model reaches, the dominant ion
producing the most emission is \ion{Fe}{8}. At the same time, some of the emission 
from the non-dominant ions does have a small impact in the synthesized
intensity of this AIA channel, depending on where and when we are
observing in the model. This can be seen when one
compares the synthetic 131~\AA{} image (panels 1-3A in Movie~1) 
and the emission coming from the dominant ions (panels 1-3B in Movie~1). 

In contrast to the previous passband, in the 131~\AA{} channel
the contribution from \ion{Fe}{8} is dominant always and everywhere (Movie~2). 
However, in a few places \ion{O}{6} and \ion{Ne}{6} have a relative
contribution that is comparable with that of the dominant ion. Other ions 
that play some role in this filter are \ion{Ne}{7} and \ion{Mg}{5},
but with relative contributions one order of magnitude smaller
than the dominant ion. Note that most of these 
ions have formation temperatures lower than the dominant ion \ion{Fe}{8}.
However, later in time (panels 3c-d in Movie~2), in addition to the 
ions mentioned, ions with formation temperatures hotter
than \ion{Fe}{8} contribute to the total intensity.  These include
\ion{Si}{12}, \ion{O}{7}, \ion{Si}{11}, and \ion{Ca}{13}, which have
contributions up to one order of magnitude smaller than the
contribution from \ion{Fe}{8} in a few locations at time 1460~s  (panels
3c-d in Movie~2).  Note that in some places the intensity signal 
appears ``noisy''. This occurs in places where the emission comes from 
transition region  plasma: in the transition region, the
temperature gradient is large, causing large jumps in temperature between
grid points so that the interpolation with $G(T,n_{\rm e})$ gives a noisy signal.

In panels 1-3B of Movie~3 we show emission in the 131~\AA{} channel from
plasma within the temperature range $\log(T)=[5.5,6]$. 
At all timesteps, most of the emission comes from temperatures  
between $5.5<\log(T)<6$ (see panels 1-3B in Movie~3). The contributions from cooler 
(panels 1-3D) or hotter (panels 1-3C) plasma is almost negligible at
all instants and almost everywhere. However, in a few small areas,
some faint emission is produced by both cooler ($\log(T) < 5.5$; see panels 1-3D) 
and hotter ($\log(T)>6$; see panels 2-3C) plasma. The 
latter is appreciable only when the 
model reaches high enough temperatures, i.e., at times 1200~s and 1460~s.

Most of the normalized intensities $I_T$ are of the order 0.025, 0.018, and
0.02 of the maximum intensity at times 880~s, 1200~s, 
and 1460~s, respectively (see the black line
in the second row in Figure~\ref{fig:inthis}). As mentioned for the 94~\AA{} channel,
at time 1200~s the emerging loop shows very strong emission. As a
result most of the normalized intensities show small values at this time. (This 
behavior is also observed in the other channels, with the exception of
the 211 and~335~\AA{}~channels, 
as discussed below). For most of the intensities, the black line is
similar to the green line, i.e.,
for the rather restricted variation of atmospheric temperatures in the model, 
we can conclude that, for any range of intensities, 
the emission in this channel comes from plasma with a temperature
range $\log(T)=[5.5,6]$. This is confirmed by the second 
row in Figure~\ref{fig:conthis} where we can observe that in general 
most of the emission (in each pixel) comes from plasma at temperatures
 $5.5<\log(T)<6$ (black lines), while only 10\% of the emission
arises from hotter plasma.   

As shown in detail in appendix~\ref{sec:abund}, for the 131~\AA{} 
channel, at all instants, the intensity calculated with  
the photospheric abundances is between 2.5 and 3.3 times smaller than 
the intensity calculated with the coronal abundances. This is caused by the
different relative abundances of O and Ne, with respect to Fe, in the 
two datasets. The contributing ions, \ion{O}{6}, and \ion{Ne}{6}/VII, 
have peak formation temperatures similar to
the dominant ion, \ion{Fe}{8}. So the temperature dependence of the
relative contribution of the different elements is less pronounced in
the 131~\AA{} channel than in the other channels. As described in 
detail in appendix~\ref{sec:elec}, the emission in this channel is not 
density sensitive. 

In summary,  at least for the temperature regimes covered by the model, 
the dominant ion producing the most emission is \ion{Fe}{8}.  The emission 
coming from other ions or/and plasma with temperatures that differ from 
the peak temperature formation of \ion{Fe}{8} is faint.

\subsubsection{171~\AA{} channel}

The SDO/AIA 171~\AA{} channel has emission from two 
dominant ions, \ion{Fe}{9} and \ion{Fe}{10}, which have 
similar formation temperatures; $7\times 10^5$~K and $10^6$~K.
As in the 94~\AA{} channel, the structures observed in the synthesized
171~\AA{} channel (panels 1-3A in Movie~1) are similar to those seen
in the emission
coming from dominant ions  (panels 1-3B in Movie~1). 
Most of the time this channel is dominated by \ion{Fe}{9}.
However, as in the 131~\AA{} channel, the importance of the
non-dominant ions varies significantly both in space 
and time. This can be observed clearly in Movie~2 which shows 
which ions contribute to the total intensity. 

The most important cool non-dominant ions 
are \ion{O}{5} and \ion{O}{6}. In a few locations, emission from
\ion{O}{5} is sometimes one order of magnitude larger  
than the emission coming from the \ion{Fe}{9} ion, with emission from
\ion{O}{6} of the same order of magnitude as the \ion{Fe}{9}
ion. However, in most locations, the emission from the dominant ions 
is the most important, by almost two orders of magnitude compared 
to the emission coming from the other ions. \ion{Fe}{10}
plays a small but still rather important role in this filter. Later in 
time, when the model is hotter, additional ions contribute as much as 
\ion{Fe}{10}, such as \ion{Ni}{14} and \ion{Ar}{10}, which have a higher 
formation temperature than \ion{Fe}{10} (see panel 3b-d in Movie~2).

In the 171~\AA{} channel, the range in temperatures where the 
$G(T,n_{\rm e})$ of the dominant ions \ion{Fe}{9} is larger than 
$1/e$ of its maximum is $\log(T)=[5.7,6.1]$. The impact of the emission coming
from plasma that is hotter (panels 1-3C in Movie~3) or cooler (panels 1-3D) 
than the range $\log(T)=[5.7,6.1]$ is similar to that observed in channel 131~\AA{}.
At all timesteps, most of the emission comes from temperatures  
within $5.7<\log(T)<6.1$ (see panels 1-3B in Movie~3) which is produced 
mostly by the dominant ions. The effects from hotter plasma
are almost negligible at all instants and almost everywhere (panels
1-3C). Even when the model reaches higher temperatures (i.e., at times
1200~s and 1460~s) these lines only provide a rather weak contribution. 
The contribution coming from plasma cooler than $\log(T)=5.7$ has only a 
limited impact (see panels 1-3D in Movie~3): it does not change the intensity 
much, nor does it change the shapes of structures seen in the synthesized 
intensity of this channel at any time. 

Most of the normalized intensity $I_T$ values are around 0.06, 0.02, and
0.1 of the maximum intensity at times 880~s, 1200~s, and 
1460~s, respectively (see the black line
in the third row in Figure~\ref{fig:inthis}). Similarly to what we found
for channel 131~\AA{}, 
the black line is overall close to the green line. 
Therefore, when the atmospheric temperatures vary within the model, 
one can conclude that, at any range of intensities, 
the emission from this channel comes 
from the temperature range $5.7<\log(T)<6.1$ (green line). 

This is confirmed by the third row in Figure~\ref{fig:conthis} where we can observe that in 
general most of the emission in each pixel comes from the range of 
temperatures  $5.7<\log(T)<6.1$ (black lines). Only slightly more than 
10\% of the emission comes from plasma cooler than $\log(T)=5.7$ and 
slightly less than 10\% of the emission comes from plasma hotter 
than $\log(T)=6.1$.  

As shown in detail in appendix~\ref{sec:abund}, the differences in emission
between the two datasets with different abundances (photospheric and coronal) 
arise from the differences in Fe abundances between both datasets.
There are also small and narrow areas where the emission is different because emission from elements other 
than Fe is substantial (e.g. \ion{O}{5}/VI, and \ion{Ne}{4}/V).

As shown in detail in appendix~\ref{sec:elec}, the emission from some structures shows 
a dependence on the plasma density, especially at later times (t=1200
and 1460~s). This is mainly due to the sensitivity to density of the strong 
\ion{Fe}{9} 171~\AA{} spectral line.
In addition, other lines which contribute less to the emission 
have also a strong dependence on density, e.g., \ion{Fe}{10} 170~\AA{}, 
\ion{O}{5} 171~\AA{}, and \ion{Ni}{14} 171~\AA{}. 

In short, our analysis indicates that the emission coming from 
other ions or/and plasma with temperatures that differ from the 
peak temperature formation of \ion{Fe}{9}/X is rather faint.

\subsubsection{193~\AA{} channel}

The dominant ion in the SDO/AIA 193~\AA{} channel is \ion{Fe}{12}
which is formed at $\sim 1.2\times 10^6$~K. However, this synthesized channel is influenced 
by the non-dominant ions in a similar manner as 
the 94~\AA{} channel. Therefore, the various structures observed 
in the synthesized 193~\AA{} emission (panels 1-3A in Movie~1) are different both in
morphology and intensity as compared to the emission coming from the dominant ion
(panels 1-3B in Movie~1).  
The contributions to the emission coming from the non-dominant
ions is rather important at all three instants, i.e., at all temperatures
 (see red areas in the panels 1-3D in Movie~1). 

The contributions of the most important non-dominant ions 
vary strongly in time and space (see Movie~2), and are mainly due to 
\ion{O}{5} and \ion{Fe}{8}, which are of the same order as \ion{Fe}{12}. 
While the relative contributions of \ion{Fe}{9} and \ion{Fe}{11} are an order of magnitude smaller than
\ion{Fe}{12} in most locations, they are significant in some
locations. In addition, a relative contribution of 
one order of magnitude smaller than that of
the most important ion(s) comes from \ion{Fe}{10}, \ion{S}{11}, \ion{Fe}{13}
and \ion{Ar}{10}. In some places the emission comes from cooler plasma 
(transition region) as can seen from the ``noisy'' signal.

In the 193~\AA{} channel, the range in temperatures where the 
$G(T,n_{\rm e})$ of the dominant ion is larger than $1/e$ of its 
maximum is $\log(T)=[6.1,6.3]$. Since the model has 
relatively low temperatures, most of the emission in this channel 
comes from plasma at temperatures lower than $\log(T)=6.1$, 
corresponding to emission from non-dominant ions which have a 
formation temperature that is lower than the dominant ion
(see panels 1-3D in Movie~3). As seen above, the 171~\AA\ shows a similar behavior. 
It is interesting to note that, even at the later times, when the model 
has sufficiently high temperatures, most of the emission comes from 
plasma that is cooler than $\log(T)=6.1$ in this channel. Note also 
that the structuring and intensities produced by the various ranges of 
temperature are completely different. This is because, at all instants,
the sources (ions) emit in different regions (compare to panels 1-3B, 
1-3C and 1-3D in Movie~3). The emission from plasma at temperatures 
higher than $\log(T)=6.3$ is negligible at all instants.
At $t=880$~s, we also see that even the emission from plasma at
temperatures  $\log(T)=[6.1,6.3]$ is faint. The emission coming from the 
range of temperature $6.1<\log(T)<6.3$ comes mostly from the dominant ions.

Most of the normalized intensity $I_T$ values are around 0.025, 0.018, and
0.033 of the maximum intensity in the box at times 880~s, 1200~s, and 1460~s, 
respectively (see the black lines in the fourth row in Figure~\ref{fig:inthis}). 
The black lines ($I_T$) are mostly a combination of the green ($I_I$) and blue ($I_L$) lines. 

The plasma with temperatures between $6.1<\log(T)<6.3$ typically shows normalized
intensity ($I_I$) values of 0.014, 0.015, and 0.02 of the maximum intensity in
the box at times 880~s, 1200~s and 1460~s, respectively (green lines). 
Plasma with temperatures lower than $\log(T)=6.1$ ($I_L$)
shows normalized intensity values of 0.016, 0.012 and 0.015 at times 880~s, 
1200~s and 1460~s, respectively (blue lines). We can thus see that the
contributions from cooler plasma change with time. 
For brighter locations, at all instants, the cold plasma (blue lines)
emits in more locations than plasma at temperatures $\log(T)=[6.1,6.3]$ (green lines).
In the fourth row in Figure~\ref{fig:conthis} one can observe that at time 880~s 
most of the emission for most of the locations arises from plasma at 
temperatures lower than $\log(T) = 6.1$ (blue lines). However, a large number of 
locations have strong contributions from plasma at temperatures $6.1<\log(T)<6.3$ 
(black lines). At time 1200~s
the dominance of the emission per pixel is shared between these two ranges of 
temperatures. Finally, at time 1460~s most of the pixels have more contributions from 
cold plasma (blue lines). For most locations roughly 30\% of its intensity comes from plasma 
with temperatures $6.1<\log(T)<6.3$ (black lines); and 20\% of its intensity comes from 
plasma hotter than $\log(T)=6.3$ (red lines). 

The assumed abundances have a significant impact on the 193\AA\
emission in two ways: 1) in large regions the ratio of the emission 
calculated using the two different abundances varies smoothly between
1 and 1.7 due to significant contributions from non-Fe ions 
throughout the field of view; 2) in some small and 
narrow features this ratio significantly
departs from 1 because the relative contribution of
emission from elements other than Fe is very substantial
(see appendix~\ref{sec:abund} for details). 

As found for the 94~\AA\  and 171~\AA\ channels, some structures 
show a dependence on electron density, because several spectral 
lines contributing to their emission are density dependent
(these lines are listed in appendix~\ref{sec:elec}).

In summary, emission coming from other ions (e.g. \ion{O}{5} and \ion{Fe}{8}) 
or/and plasma with temperatures that are cooler than the peak 
temperature formation of \ion{Fe}{12} contribute considerably to the total
emission.

\subsubsection{211~\AA{} channel}

The SDO/AIA 211~\AA{} channel has emission from one dominant ion,
\ion{Fe}{14}, which has a peak formation temperature of
$\approx 2\times 10^6$~MK, i.e.\
near the upper limit of the temperature range of our model.
We can therefore expect that non-dominant ions will play a larger role than in
atmospheres with higher average temperatures. Figure~\ref{fig:aia211} illustrates
that most of the synthesized emission of channel 211~\AA{} 
(panels 1-3A) comes from the so-called non-dominant ions 
(panels 1-3C). The emerging loop seen in channel 94~\AA{} at time
1200~s is also visible in this channel. However, in channel 
211~\AA{} this structure in particular, as well as most of the other
small structures observed, shows a markedly different shape than that
seen in the emission coming from the dominant ion (panel 2B). 
This is because this channel has strong contributions 
from ions with cooler formation temperatures (see panels 1-3C and 1-3D in 
Figure~\ref{fig:aia211} and the corresponding images in Movie~2). 

In a similar manner to channel 94~\AA{}, the ions which 
provide the most important contributions differ from gridpoint 
to gridpoint due to the dynamic atmosphere (see Movie~2). 
However, at all instants, all the non-dominant ions significantly 
contributing to the emission are formed at temperatures lower 
than the peak formation temperature of the dominant ion. This is 
presumably because the temperature of the model barely reaches 
temperatures above the peak formation temperature of the dominant ion.
The most important ions in terms of emission are \ion{O}{5} and
\ion{O}{4}. However, there are other ions that are important
contributors as well, depending on the timestep and position 
in the box. These contributions are due the ions \ion{Ne}{5}, 
\ion{Ne}{4} and \ion{N}{5}, \ion{Fe}{12} and \ion{Fe}{13} (in 
addition to small contributions from the ``dominant'' ion \ion{Fe}{14}).

In the right panels in Movie~2 we see additional ions which 
also contribute, such as \ion{Fe}{8}, \ion{Si}{8}, \ion{Ni}{11}, 
and \ion{Fe}{10}. At $t= 880$~s, there are a large variety of 
contributions coming from various ions. Later in time, at
$t=1200$~s, when the coronal plasma is hotter, more integrated 
columns have a clear dominance of  contributions coming from 
the ions \ion{Fe}{14} and \ion{Fe}{13}. This is even more 
pronounced at $t=1460$~s, where we find a clear dominance
by the dominant ion, though even here, not at all locations.
The highly dynamic stratification is thus variable enough to 
considerably change the importance of the emission from each 
contributing ion. 

The range in temperatures where the $G(T,n_{\rm e})$ of the dominant
ion is larger than 1/e of its maximum is $\log(T)=[6.2,6.4]$ for the
211~\AA{} channel (see panels 1-3B in Figure~\ref{fig:temp}).
The impact of the emission coming from hotter or cooler 
plasma than this range is similar to the ones observed in 
channel 193~\AA{}. Therefore, since the model is relatively cold, 
most of the emission comes from temperatures lower than $\log(T)=6.2$, 
and its origin is from the non-dominant ions which have a formation 
temperature lower than that of the dominant ion. This happens at all instants, 
but when the temperature of the model is high enough, some structures 
formed at temperatures between $6.2<\log(T)<6.4$ are also observed in the 
synthesized intensity of the channel.
In a similar manner as in channel 193~\AA{}, it is important 
to note that the shapes of structures and intensities coming from the various 
temperature ranges are completely different. This is because the sources and 
regions responsible for the various temperature ranges are physically
distinct. The emission coming from plasma at temperatures larger 
than $\log(T)=6.4$ is negligible. At $t=880$~s
the emission coming from plasma with temperatures for
which $\log(T)=[6.2,6.4]$ is also faint.

In general, the emission from plasma in the range of
temperatures  $6.2<\log(T)<6.4$ comes mostly from the dominant ion. 
However, one can see some structures which are slightly 
larger or only appear in the intensity emitted by plasma with
temperatures  $6.2<\log(T)<6.4$. This is clearly visible when
comparing the emission coming from this range (Movie~3) with the 
intensity coming from the dominant ion in Movie~1. These differences
arise because the channel has some contributions from the \ion{Fe}{13} which 
has a formation temperature relatively close to the dominant ion.

Most of the normalized intensity $I_T$ values are around 0.02, 0.02, and
0.07 of the maximum intensity at times 880~s, 1200~s, and 1460~s, 
respectively (see the black lines in the fourth row in Figure~\ref{fig:inthis}).
Note that, in comparison with other channels, 
at time 1200~s the emission from the emerging loop (relative to the
rest of the box) is not as strong as in the other channels. 
At time 880~s, at any value of intensity, the black line ($I_T$) is the same
as the blue line ($I_L$). This means that all the emission comes from cold 
plasma (blue lines, $I_L$). 
However, later in time, some emission comes from the range of
temperatures  $\log(T)=[6.2,6.4]$ (green lines, $I_I$). This becomes
clear when viewing the histogram: most of the normalized intensity
values emitted by plasma with temperatures $6.2<\log(T)<6.4$ ($I_I$) are 0.0, 
0.016, and 0.04 at times 880~s, 1200~s and 1460~s, respectively (green lines). 
The normalized intensity values emitted by plasma with temperatures lower than 
$\log(T)=6.2$ are 0.02, 0.013 and 0.016 of the maximum of the box at times 880~s, 
1200~s and 1460~s, respectively (blue lines, $I_L$). Again we see that the
contributions of cold and hot plasma to the intensity in each pixel vary 
at times 1200~s and 1460~s. It is interesting to see that 
the intensity histogram from the cold plasma (blue lines) 
is larger below intensities 0.04 and 
above intensities 0.1 than the histogram of the intensity emitted
by plasma with temperatures  $6.2<\log(T)<6.4$ (green lines). 
Therefore, most of the emission of the weak background comes
from cold plasma, but also the strong intensity features come from cold plasma (blue lines). 

In the fourth row in Figure~\ref{fig:conthis} one can observe that at time 880~s, 
for all the integrated columns,
all the emission comes from temperatures lower than $\log(T) = 6.2$ (blue lines).
At time 1200~s, the dominance of the emission per point is shared between temperatures lower than 
$\log(T)=6.2$ (blue lines) and the range of temperatures $6.2<\log(T)<6.4$ (black lines).
Finally, at time 1460~s most of the integrated columns have more contribution from 
plasma with temperature  $6.2<\log(T)<6.4$ (black lines), and for most of them 
roughly 30\% of its intensity is emitted by cold plasma (blue lines). 

As found for  the 193~\AA\ channel, the difference of the synthetic
images assuming photospheric or coronal abundances is ``twofold'':
1. in large regions the ratio of the emission calculated using the two different
sets of abundances varies smoothly; 2. there are small and 
narrow features where this ratio significantly
departs from 1 (see appendix~\ref{sec:abund} for details). 

Similarly to the 94~\AA\, 171~\AA\, and 193~\AA\ channels some structures 
show a dependence on density. Some spectral 
lines contribute to the emission and are strongly density dependent, 
i.e., \ion{O}{5} 215~\AA{}, \ion{Fe}{14} 211~\AA{}, \ion{O}{4} 208~\AA{}, 
\ion{Fe}{13} 202~\AA{}, \ion{Fe}{12} 211~\AA{}, and \ion{Fe}{10} 207~\AA{}
(see appendix~\ref{sec:elec}).

In short, emission coming from other ions 
(e.g. \ion{Fe}{8}, \ion{Si}{8}, \ion{Ni}{11},  and \ion{Fe}{10}.) 
or/and plasma with temperatures that are cooler than the peak 
temperature formation of \ion{Fe}{14} have strong contributions to 
the total emission. Such contribution is dominating everywhere when the 
model is cooler (at $t=880$~s, $Max(T)=1.6\,10^{6}$~K, see 
Figure~\ref{fig:tempdist}).

\subsubsection{304~\AA{} channel}

The dominant ion in the SDO/AIA 304~\AA{} channel is \ion{He}{2}. 
However, the line formation of \ion{He}{2} is poorly understood, so it is 
not clear that the optically thin approach is valid \citep{Feldman:2010lr}. 
Ignoring this issue and assuming optically thin radiation, the various structures observed  
in the synthesized 304~\AA{} channel (panels 1-3A in Movie~1) are very similar 
to the emission coming from the dominant ion (panels 1-3B in Movie~1). 
In the few places where the intensity suffers some small impact from the
non-dominant ions, it strongly depends on the location and timestep of the model (see 
panels 1-3C and 1-3D). 

In general, the emission coming from non-dominant ions is negligible. 
However, \ion{Si}{11} produces constant, though weak, emission in
several regions, as is clear from panels 1-3d in Movie~2. 
This \ion{Si}{11} contribution is more important when the model gets hotter,
and, in some faint locations, its emission is more than two
times larger than that of the dominant ion (see panels 2a-d and 3a-d in
Movie~2). In a few locations, other non-dominant ions play a small role, although
these are highly dependent on time and space. In these locations, they
emit up to one magnitude less than the dominant ion. These ions have
formation temperatures higher than that of the dominant ion and
include: \ion{O}{4},  \ion{O}{3}, \ion{Si}{9}, \ion{Mg}{8}, \ion{Si}{8},  
\ion{Fe}{13}, \ion{Fe}{15} and/or \ion{S}{12}. 

For the 304~\AA{} channel, the temperature range where the $G(T,n_{\rm e})$ 
of the dominant ion is larger than $1/e$ of its maximum is $\log(T)=[4.8,5.1]$.  
The impact of the emission coming from hotter (panels 1-3C in Movie~3) 
or cooler plasma (panels 1-3D) than the range $4.8<\log(T)<5.1$ is similar 
to the observed in channels 131~\AA{} and 171~\AA{}. Therefore, at all three 
instants, most of the emission comes from temperatures  between 
$4.8<\log(T)<5.1$. The emission from plasma that is hotter than $\log(T)=5.1$ 
and plasma that is cooler than $\log(T)=4.8$ is almost negligible at all 
instants and almost everywhere. There are some faint contributions to 
the total intensity in small regions, both from cooler and hotter plasma, 
but they do not change the shape of the structure observed in the 
synthetic 304~\AA{}. The contributions of the various temperature range do 
not change with time. The emission from plasma at temperatures  
$4.8<\log(T)<5.1$ comes mostly from the dominant ion.

Most of the normalized intensity values are of the order of 0.01 of
the maximum intensity at any time (see the black lines 
in the sixth row in Figure~\ref{fig:inthis}). Therefore most of the box shows 
weak emission, with a few very bright points. When the intensity is
roughly larger than 0.015 of the maximum intensity in the box, 
the emission comes from plasma
with temperatures  $\log(T)=[4.8,5.1]$ (green lines).  
However, the rather faint locations (normalized intensity below 0.1) 
have important contributions from hot plasma (red lines). 
In fact, in the sixth row in Figure~\ref{fig:conthis} one can observe that in general 
most of the locations have emission emitted by hot plasma (red lines). 
It is only at $t=880$~s when most of the relative contributions 
comes from temperatures within the range of temperatures
$4.8<\log(T)<5.1$ (black lines).   

The ratio of the intensities calculated assuming the two different
sets of abundances ranges from 0.4 to 1 and the values significantly lower than 1 are 
due to the contribution of \ion{Mg}{8} and \ion{Si}{8}/IX/XI to the 
emission observed in this channel (see appendix~\ref{sec:abund}). 
The 304~\AA{} channel has faint contributions from a few spectral lines
that are strongly density dependent: \ion{Si}{8} 316~\AA{}, and 
\ion{N}{4} 332~\AA{} (see appendix~\ref{sec:elec}). 

We find that in the 304~\AA\ channel the emission appears dominated by
\ion{He}{2} and the 
contribution from other ions is faint. However, we have to keep in mind that our
calculations assume the optically thin approximation which might not be valid for 
\ion{He}{2} 304~\AA\ line formation which is poorly understood.

\subsubsection{335~\AA{} Channel}

The dominant ion in the 335~\AA{} SDO/AIA passband is
\ion{Fe}{16}, with peak formation temperature $\approx 2.5\times 10^6$~K.
The synthetic emission in this channel appears to be influenced by the
non-dominant ions in a similar manner as the 211~\AA{} channel. 
The formation temperature of \ion{Fe}{16} is higher than the
temperatures reached by our model, and this is one of the reasons why
most of the 335\AA\ emission (panels 1-3A in Movie~1) comes from the
non-dominant ions (panels 1-3C in Movie~1). At $t=880$~s, the emission
only comes from
the non-dominant ions because the temperature of the model is too low 
to produce significant \ion{Fe}{16} emission (see panels 1A-D in Movie~1).
However, later in time (see at $t= 1200$~s panels 2A-D and at $t=1460$~s panels 3A-D) 
some emission stemming from \ion{Fe}{16} (panels 2B and 3B) appears.

The synthetic emission is produced by a large number of non-dominant
ions, all of them cooler than the dominant ion (see Movie~2). The
relative contributions of the non-dominant ions change considerably 
as a result of the dynamics of the atmosphere. At $t=880$~s, when the model has low 
temperatures, \ion{O}{3}, \ion{He}{2}, \ion{Mg}{8}, \ion{Fe}{8} and
\ion{O}{6} produce the strongest emission. 
Other ions have strong contributions, such as \ion{N}{4}, 
\ion{Si}{10}, \ion{Fe}{11}, \ion{Ne}{6} and/or \ion{Fe}{10}, 
amongst others. 
At $t=1200$~s, when the model has higher temperatures, the relative
contribution from any of the ions \ion{O}{3}, \ion{He}{2},
\ion{Fe}{14}, \ion{Fe}{16}, \ion{Fe}{8} and \ion{O}{6} could be the
most important. In addition, \ion{N}{4}, 
\ion{Mg}{8}, \ion{Al}{10}, \ion{Ne}{6} and/or \ion{Ca}{7} 
have strong contributions at some locations.
Finally, at $t=1460$~s, when the model reaches the highest  temperatures, 
the most important emission comes from \ion{Fe}{16} in most places,
but there are still locations with important contributions from  
\ion{O}{3}, \ion{He}{2}, \ion{Fe}{14}, \ion{Fe}{8} and \ion{O}{6}.
In addition, as in the previous instants, \ion{N}{4}, 
\ion{Mg}{8}, \ion{Al}{10}, \ion{Ne}{6} and/or \ion{Ca}{7} 
also have strong contributions in a few locations.

The temperature range where the $G(T,n_{\rm e})$ of the dominant
ion is larger than $1/e$ of its maximum is $\log(T)=[6.3,6.7]$ for the
335~\AA{} channel. The impact of the emission coming
from hotter or cooler plasma than the range $6.3<\log(T)<6.7$ 
is similar to that observed in channels 193~\AA{} and 211~\AA{}. Therefore, 
since the model has rather low temperatures, most of the emission
comes from temperatures lower than $\log(T)=6.3$. This happens at all instants, 
but when the temperature of the model is high enough, some faint structures 
observed at temperatures between $6.3<\log(T)<6.7$ can be appreciated 
in the synthesized intensity of the channel which comes mostly from the 
dominant ion. In a similar manner as observed in channels 193~\AA{} 
and 211~\AA{}, it is also important to note that the structures and intensities 
coming from the various temperature ranges are completely different. 

Similarly to the 171~\AA{} channel, most of the contributions at
temperatures that are lower than $\log(T)=6.3$ come from non-dominant
ions. All of these ions have a formation temperature that is lower than 
the dominant ion. 

Most of the normalized intensity values are around 0.025, 0.02, and
0.06 of the maximum intensity at times 880~s, 1200~s, and 1460~s, 
respectively (see the black lines in the last row in Figure~\ref{fig:inthis}).
At time 880~s, and 1200~s at any range of intensities, the black
line is almost the same as the blue line. This means that all the
emission comes from cold plasma. 
However, later in time, i.e., at time 1460~s, a large amount of the weak emission 
comes from plasma with temperatures  $\log(T)=[6.3,6.7]$. For the
earlier times, most of the normalized intensity values from temperatures between $6.3<\log(T)<6.7$ are negligible. It is about 0.025 of the maximum intensity at time 1460~s (green lines). 
Most of the intensity coming from plasma at temperatures lower than 
$\log(T)=6.3$ is 0.025, 0.02 and 0.055 of the maximum at time 880~s, 
1200~s and 1460~s, respectively (blue lines). There is once again a
variation with time of the relative brightness of the emission from
plasma in the various temperature bands. 

In the last row in Figure~\ref{fig:conthis} one can observe that at time 880~s and 
1200~s for all integrated columns all the emission comes from temperatures 
lower than $\log(T) = 6.3$ (blue lines). At time 1460~s
most of the emission per integrated column for most of them comes from cold
plasma (blue lines), with about 30\% contribution from plasma with a
temperature within $6.3<\log(T)<6.7$  (black lines).

As found in the 193 and 211~\AA\ channels, in the 335~\AA{} 
channel the difference of the synthetic images assuming photospheric 
or coronal abundances is ``twofold'' (see appendix~\ref{sec:abund} for details).
We also conclude that the plasma density does not affect the 335\AA\ emission 
(see appendix~\ref{sec:elec}). 

In summary, for the temperature distributions of our model, which barely reach the 
temperature of peak formation of the dominant ion 
\ion{Fe}{16}  ($\sim 2.5$~MK), the
emission is predominantly coming from ions which have a peak temperature
formation lower than \ion{Fe}{16}, such as \ion{O}{3}, \ion{He}{2}, \ion{Mg}{8}, \ion{Fe}{8} and
\ion{O}{6}. 

\section{Conclusions}
\label{sec:concl}

We  have investigated the influence of non-dominant ions on, and 
the temperature dependence of, the narrow wavelength bands of the 
various SDO/AIA channels. In order to carry out this study, 
emission from the different channels has been synthesized based on a 
realistic 3D simulation. The dynamics
arising from the evolution of the model allowed us 
to investigate the origin of the emission for a large variety of
stratifications that occur naturally in the evolving atmosphere. 
In addition, the increasing coronal temperatures
(with time) in the simulated box allow us to study the impact of
the coronal temperature on what is observed with AIA.
Because of the large number of AIA channels and simulated snapshots,
most of the results are made available as online movies.

In general, the large variety of stratifications 
produces a considerable variation of the origin of the emission in the
various AIA passbands. It is found that non-dominant ions can
dominate the AIA channels, depending on the location and
the time. As a consequence, we find that the dominant
emission in AIA pixels can often arise from plasma that is emitting at
temperatures that are quite different from what one would expect based on
typical instrument descriptions.

In summary, we observed that the 131, 171, and 304~\AA{} AIA passbands 
are the least influenced by these effects. Almost everywhere, 
and at all times (i.e., time and/or temperature of the corona given the
increasing temperatures in the box), most of the emission in these
channels comes from the dominant ion(s). Note that the 131~\AA{}
and 171~\AA{} channels have two dominant ions. In the former channel
the most important ion (in our calculations) is \ion{Fe}{8} because the 
temperature range of our model is too low to produce \ion{Fe}{20} 
emission; however,  \ion{Fe}{20} is expected to be significant only for 
flaring plasmas.
In contrast, in the 171~\AA{} channel, the 
dominant spectral lines come from ions with formation temperatures that are rather similar,
i.e., \ion{Fe}{9} and \ion{Fe}{10} (although the most important is
\ion{Fe}{9}). In all three of these channels the contribution from 
non-dominant ions is significant only in a few places at most, and
even then it is rather small. In addition, most of the
emission comes from plasma with temperatures near the formation temperature of the 
dominant ion(s). The 304\AA\ results are less solid since
the line formation of \ion{He}{2} is poorly understood (so that 
it is not clear that the optically thin approach is valid).

On the other hand, and from better to worse, the origins of the emission in the 94, 193, 211
and 335~\AA{} passbands are strongly influenced by emission from non-dominant ions
or structures with non-typical temperatures. In different regions, 
and at different times in the simulation, varying (but significant)
fractions of the emission in these AIA channels is
emitted by the so-called non-dominant ions. Channel 211 and 335~\AA{}
have a dominant ion with peak formation temperature close to the
higher temperatures reached by the modeled corona. 
Therefore, most of the emission for these two channels 
comes from spectral lines of ions with a formation 
temperature that is lower than that of the dominant ion: the emission comes from cooler plasma.  
This is more clear at the first timestep of the simulation because the
model is cooler at that time. Note that 
the 94~\AA{} channel has two dominant ions \ion{Fe}{10} and \ion{Fe}{18}, and the one that is the most
important here is 
\ion{Fe}{10}, again because of the relatively low temperatures in the
model. 
In all of these channels (94, 193, 211 and 335~\AA{}),
significant emission comes from non-dominant ions, at least in some
places. Because we do not know, a priori, what the real temperature
distribution on the sun is, it is clear that the interpretation of AIA
data in these passbands can be influenced strongly by emission from 
these non-dominant ions. As a consequence, the emission might come from plasma 
with temperatures that are (usually) lower than the formation temperature of the 
dominant ion. Given the temperatures reached in our model, the
warnings above should especially be taken into account
when observing quiet Sun and coronal hole regions. 

We have degraded the synthesized images to the SDO/AIA resolution 
and found that the temperature sensitivity and influence of the
non-dominant ions is not considerably affected 
by the spatial resolution of the observations. Therefore, the results described above for the full resolution
of the model can be directly applied to observations at SDO/AIA resolution.

We have also calculated the synthetic intensities for both coronal and
photospheric abundances. The different assumptions about abundances can
have significant effects on what is observed in the AIA passbands. The 
main differences are caused by the contributions from ions (e.g., O ions) which have 
a different ratio between photospheric and coronal abundances than
that of the dominant ions (mostly Fe ions).

We have investigated the effect of the density
sensitivity of the contribution function ($G(T,n_{\rm e})$, instead of the
usually assumed $G(T)$). The channels that
show the largest differences using $G(T,n_{\rm e})$ instead of $G(T)$ are the 
94, 171, 193, 211 and 304~\AA{} passbands.

We note that our calculations assume the validity of ionization equilibrium
conditions, and are based on the atomic and spectral line parameters
in the CHIANTI database. While both assumptions are commonly used, it
is clear our conclusions can be significantly changed if either
non-equilibrium ionization plays a significant role, or if the CHIANTI
database does not accurately describe the observed wavelength ranges. We
know, for example, that the latter issue is important for the 94 and
131~\AA{} passbands. There is now observational evidence that
both of these passbands contain some
spectral lines that are not included in CHIANTI
\citep{Testa:2011fk,Schmelz:2011fj,Aschwanden:2011fj}, and therefore
are not accounted for in our analysis. 

Nevertheless, within the limitations of our assumptions, we conclude 
that there are only a few channels where the
dominant emission is as expected from the canonical values. This means
that the interpretation of AIA data, especially in quiet Sun and
coronal holes is fraught with ambiguity, and if one depends on the
canonical values for temperature and/or ions one can easily be led to the
wrong conclusions. 

There are several possibilities to mitigate this issue. 
One possibility is to study the spatial and temporal differences in
emission from different passbands to estimate how likely it is that some signals
are caused by ``contaminating'' lines. This is the approach taken by 
\citet{De-Pontieu:2011lr} when investigating the coronal counterparts
to spicules. These authors exploit the fact that the observed
(simultaneous) signals in 171 and 211~\AA{} are spatially and
temporally offset to argue that a coronal counterpart to spicules does
exist, and that a contaminating signal from cooler ions is
highly unlikely responsible for the observed signals. As shown in the above, the contaminating
spectral lines in these passbands are from identical or similar
ions (\ion{O}{5} and \ion{O}{6}) and would show much stronger similarity
in time and space than observed \citep[contradicting the claims by][]{Madjarska:2011uq}.
We postpone a detailed investigation of this promising method to a
future paper.

Another, more general approach is to perform a DEM reconstruction 
using many AIA channels. However, we will describe in a follow up paper 
the effects of the temperature and density stratification in the solar
atmosphere on obtaining reliable DEMs.
The overlap in the line-of-sight of coronal structures with a wide range of
temperatures and densities can render 
some of the assumptions underlying DEM reconstructions invalid.
A more promising avenue for analyzing and interpreting the SDO/AIA
observations may be to compare to such observations with synthetic
observables from forward models such as the one used in the current
work.

\section{Acknowledgments}

This research has been supported by a Marie Curie Early Stage
Research Training Fellowship of the European Community's Sixth Framework
Programme under contract number MEST-CT-2005-020395: The USO-SP 
International School for Solar Physics. 
Financial support by the European Commission through the SOLAIRE Network 
(MTRN-CT-2006-035484) and by the Spanish Ministry of Research and Innovation 
through project AYA2007-66502 is gratefully acknowledged. 
B.D.P. is supported by NASA grants NNX08AL22G and NNX08BA99G and NASA
contract NNM07AA01C (Hinode).
P.T. was supported by contract SP02H1701R from Lockheed-Martin to the
Smithsonian Astrophysical Observatory.
Hinode is a Japanese mission developed 
by ISAS/JAXA, with NAOJ as
domestic partner and NASA and STFC (UK) as international partners. It
is operated in cooperation with ESA and NSC (Norway).
 The 3D simulations have been run with the Njord and Stallo cluster from the Notur
project and the Pleiades cluster through computing grants SMD-07-0434, SMD-08-0743,
 SMD-09-1128, SMD-09-1336, SMD-10-1622 and SMD-10-1869 from the High
 End Computing (HEC) division of NASA. 
We thankfully acknowledge the computer and supercomputer 
resources of the Research Council of Norway through grant 170935/V30 and through 
grants of computing time from the Programme for Supercomputing. 
To analyze the data we have used IDL and Vapor (http://www.vapor.ucar.edu).

\bibliographystyle{aa}

\bibliography{aamnemonic,collectionbib}

\appendix

\section{Instrumental contributions}
\label{sec:reso}

\begin{figure}
  \includegraphics[width=0.95\textwidth]{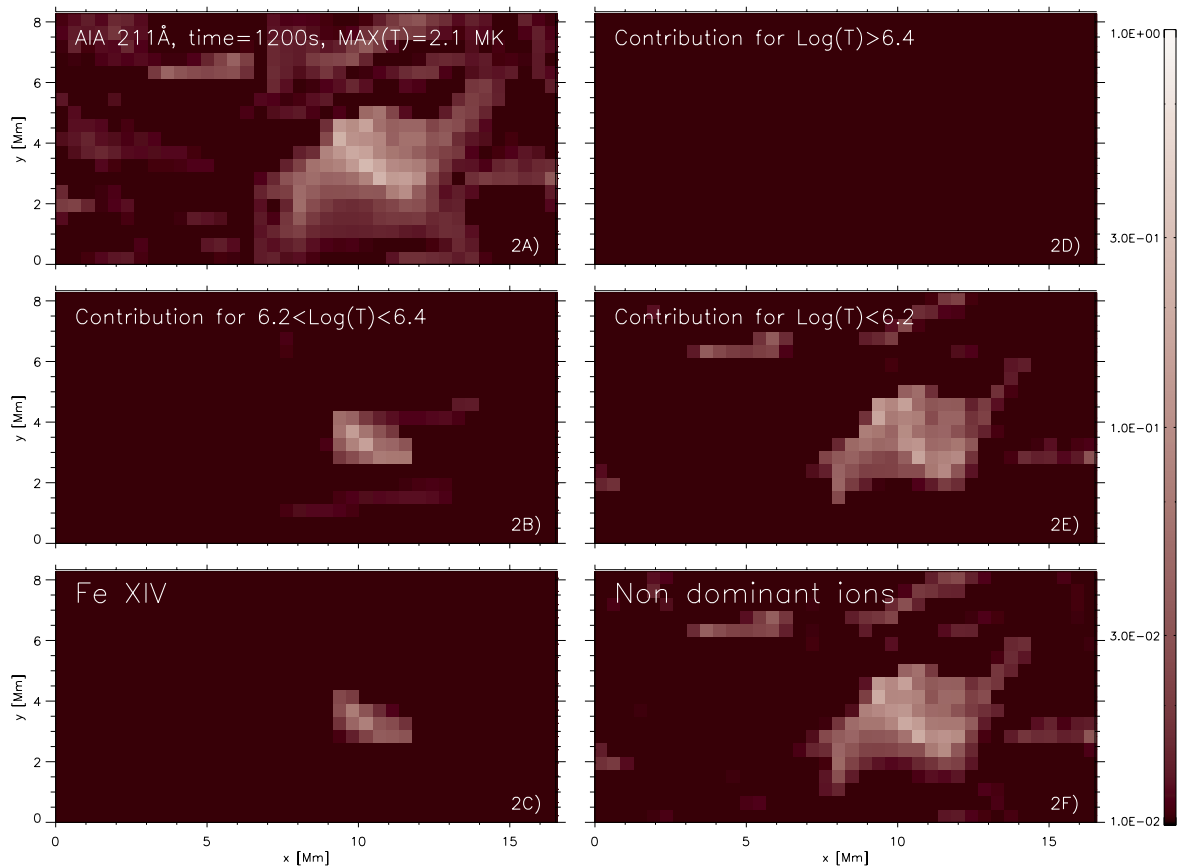}
   \caption{\label{fig:pixel} The synthetic 211\AA\ image, at the
     AIA spatial resolution, is shown in panel 2A. 
     The  intensities $I_I$, $I_D$, $I_H$,$I_L$, $I_{ND}$ are shown in  
   panel 2B, 2C, 2D, 2E, and 2F respectively. The color-scheme is shown in the right 
   colorbar in logarithmic scale and all panels are normalized to the maximum of 
   the synthesized intensity (panel 2A). All panels have the SDO/AIA spatial resolution.}
\end{figure}

The presence of small-scale structures in the emergent intensity
images from our simulations, and their impact on the importance of
non-dominant ions, indicates that the relatively coarse spatial
resolution of AIA images could change some of our results concerning the extent to which
non-dominant ions and different temperature ranges
contribute to the emission in each channel (Section~\ref{sec:diffion}). 
In order to asses this effect, we degraded the
synthesized intensities to the AIA spatial resolution.  
Figure~\ref{fig:pixel} and Movie~4 show the same plots as shown in 
Figures~\ref{fig:aia211} and~\ref{fig:temp}, and Movies~1 and~3 but at
the AIA spatial resolution. In the same 
manner as in the previous sections we show here one 
example (Figure~\ref{fig:pixel}), but the online Movie~4 shows the
same panels for all channels and at the three different instants ($t=[880,1200,1460]$~s). 
In Figure~\ref{fig:pixel}, the synthesized 211~\AA{} image at the SDO/AIA 
spatial resolution at time 1200~s is shown in the panel labeled 2A. 
The emission from the temperatures between $6.2<\log(T)<6.4$ ($I_I$) is
shown in panel 2B. The emission from the dominant ion (\ion{Fe}{14},
$I_D$) is shown in panel 2C. The emission from temperatures higher 
than $\log(T)=6.4$ for the 211~\AA{} channel ($I_H$) is shown in panel 2D. The emission
from temperatures smaller than $\log(T)=6.2$ ($I_L$) is shown in panels 2E. 
Finally, the emission from the non-dominant ions ($I_{ND}$) 
is shown in the panel 2F. All of these images are at the SDO/AIA
spatial resolution. See Movie~4 for all the channels and all three 
instants, the layout is the same as in the figure.  

In general, the detailed descriptions given in section~\ref{sec:diffion} 
are also valid when the synthesized intensities are 
convolved and pixelized to the SDO/AIA resolution. However, the
smaller structures disappear or are mixed with emission coming from
other structures.  

\subsection{94, 193 and 211~\AA{} channels}

The 94, 193, 211~\AA{} channels show strong contributions 
from so-called non-dominant ions ($I_{ND}$, panels 1-3F in Movie~4): most features 
seen in the emission from the dominant ion(s) ($I_D$, panels 1-3C in 
Movie~4) are different, both in shape and intensity, compared to
the synthesized total intensity (panels 1-3A in Movie~4). As mentioned in 
Section~\ref{sec:diffion}, most of the emission
comes from plasma cooler than the formation temperatures of the 
dominant ion ($I_L$, see panels 1-3B). It is also important 
to note that the structures seen in the intensity images 
at the SDO/AIA resolution appear very different for the various  
temperature ranges (compare to panels 1-3B, 1-3D and 1-3E). 
The reason for this is that, for these channels, the source of the 
emission for the various temperature ranges 
comes from physically distinct features or regions. This is also noted
and described in the previous sections and is clearly seen in the full resolution 
Movies~1 and~3.

\subsection{131, 171 and 304~\AA{} channels}

When taking into account the SDO/AIA resolution, most of the emission for the
131~\AA{}, 171~\AA{}, and 304~\AA{} channels comes from the dominant ion(s) 
($I_D$, panels 1-3C in Movie~4), and in only a few places do we find
that the intensity changes ($I_T$) due to a small contribution coming 
from the non-dominant ions ($I_{ND}$, see panels 1-3F in Movie~4). 
Therefore, at all three instants, most of the emission comes from temperatures  
near the formation temperature of the dominant ion(s) ($I_L$, see panels 1-3B in Movie~4). 
The emission from the hot  ($I_H$, see panels 1-3D) and the cold 
plasma  ($I_L$, see panels 1-3E) are almost negligible  
at all instants and almost everywhere. Taking into account 
the SDO/AIA resolution, the temperature dependence and the 
contribution of the different ions, we find that the results discussed in 
Section~\ref{sec:diffion} do not change considerably. 

\subsection{335~\AA{} channel}

As mentioned previously, the model does not reach temperatures
representative of strong active regions or flares. Most of the
emission in this channel is thus from lines coming from the non-dominant
ions that form at temperatures lower than the dominant ion ($I_{ND}$,
see panels 1-3F in Movie~4). Therefore, most of the emission
comes from plasma at temperatures lower than the formation temperature of the dominant ion 
($I_L$, see panels 1-3E in Movie~4). Once again, the results at the SDO/AIA spatial resolution
do not differ significantly from the description given in Section~\ref{sec:diffion} 
which are at the full model resolution. 

\section{Abundances: coronal or photospheric?}
\label{sec:abund}

The intensity of the various strong lines in each AIA
passband can be significantly affected by the assumed values for the
element abundances, which are in turn also important to interpret
the observational data. In order to investigate the effect of the chemical 
composition we synthesized AIA observations for two different sets of 
abundances: photospheric abundances \citep{Grevesse:1998uq}, 
and coronal abundances \citep{Feldman:1992qy}. 
Element abundances in solar coronal plasmas often depart from their
photospheric values, and these abundance anomalies appear
to correlate with the element's First Ionization Potential (FIP). 
Low FIP elements ($< 10$~eV; e.g., Fe, Mg, Si) are found to be enhanced in
the corona typically by a factor $\sim 3-4$, and high FIP elements
($\gtrsim 10$~eV; e.g., O, Ne) have coronal abundances close to
their photospheric values (see e.g., reviews by
\citealt{Meyer:1985lr,Feldman:1992qy}).  
A set of ``coronal abundances'' compiled by \citep{Feldman:1992qy} is
often used for studies of coronal emission - and is assumed in the
calculations for the standard temperature responses of the AIA EUV
channels - although detailed studies of the chemical composition of
solar plasmas have shown that this ``FIP effect'' varies in different
types of solar features (with, e.g., coronal holes, fast solar wind,
and newly emerged active regions showing abundances close to
photospheric), from structure to structure, and in time (see e.g.,
\citealt{Testa:2010fk} for a brief review). 
 
\begin{figure}
  \includegraphics[width=0.7\textwidth]{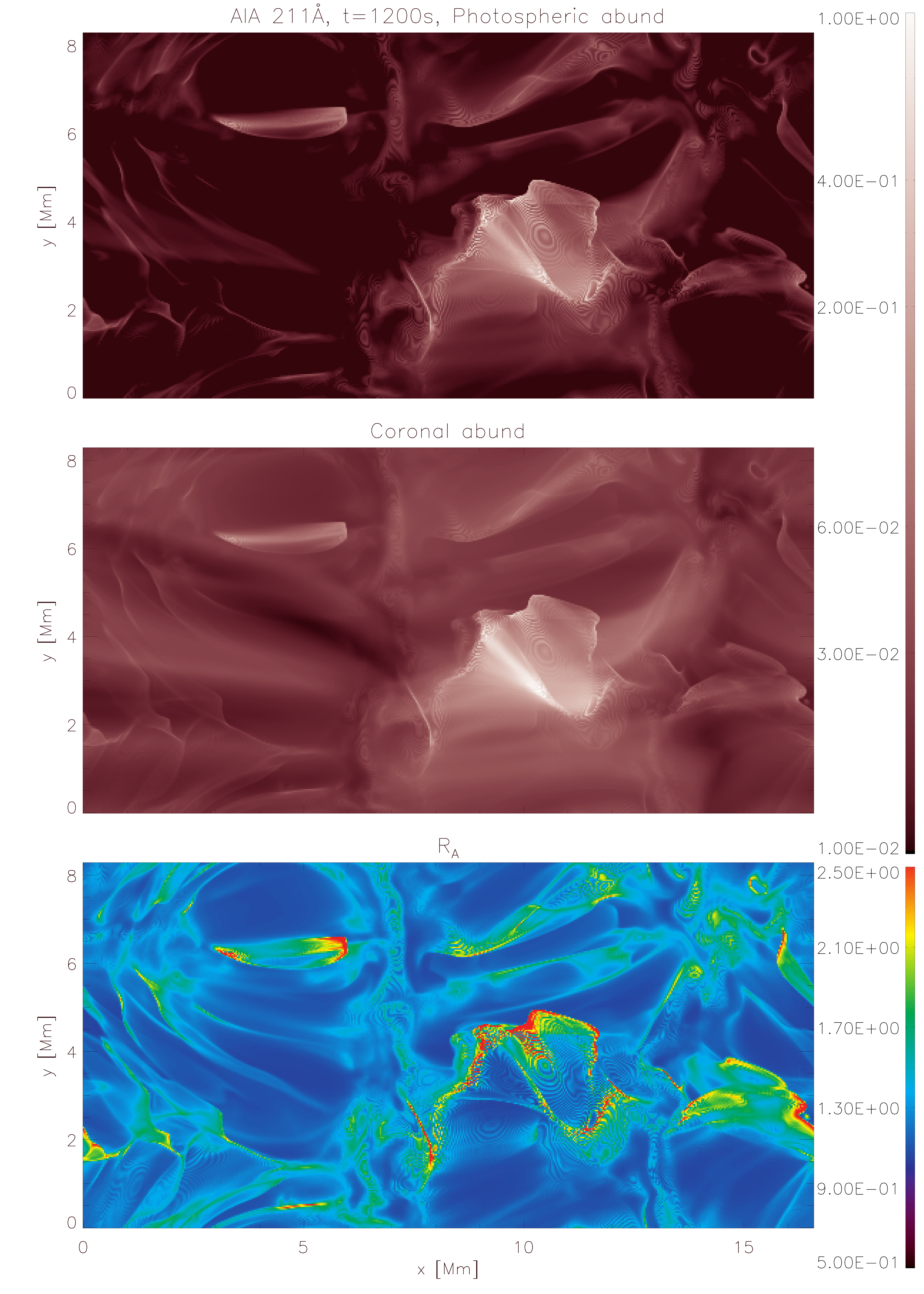}
 \caption{\label{fig:abund} A comparison of the 211~\AA\ intensity, at
   time 1200~s, using photospheric (top panel) and coronal (middle
   panel) abundances. The color-scheme is shown in the top-right
   colorbar using logarithmic scale.  The $R_A$ term (see
   Equation~\ref{eq:abund}) is shown in the bottom panel. The
   color-scheme is shown in the bottom-right colorbar.} 
\end{figure}

As in the previous sections, we here show one example (the 211\AA{}
channel; Figure~\ref{fig:abund}), and refer the reader to the online
Movie~5 which shows the results for all the channels, and at the
three different instants ($t=[880,1200,1460]$~s). 
Figure~\ref{fig:abund} shows the synthesized 211~\AA{} image at 
time 1200~s using photospheric abundances (top panel) or coronal
abundances (middle panel). The bottom panel shows the term
$R_A$ which is defined as:

\begin{eqnarray}
R_A=\left(\frac{I_{TPh}}{I_{TCo}}\right)\left(\frac{A_{DCo}}{A_{DPh}} \label{eq:abund}\right)
\end{eqnarray}

\noindent where $A_{DPh}$, and $A_{DCo}$ are the photospheric abundance, and 
the coronal abundance of the dominant ion (see last column in
Table~\ref{tab:chan}). 
$R_A$ is therefore close to 1 where the dominant ion really dominates
the emission. $R_A$ is significantly different from 1 where there is
significant contribution to the emission from non-dominant ions of 
elements that have relative abundances in the two datasets different from
the element emitting the dominant line.
We note that, for all the AIA EUV channels, the dominant line(s) are
produced by \ion{Fe}{0} ions, at various ionization stages, except for the 304~\AA\
channel which is dominated by emission from \ion{He}{2}. In the previous sections we 
used the photospheric abundance \citep{Grevesse:1998uq}. 

\subsection{94, 171~\AA{} channels}

For the 94, and 171~\AA{} channels the synthetic
intensity calculated with the photospheric abundances is, in most
areas and at all instants, about 3.8 times smaller than the intensity
calculated with the coronal abundances. $R_A$ is close to 1 (dark
blue color in Figure~\ref{fig:abund} and Movie~5), reflecting the
differences between the two datasets in the abundance of Fe, which
generally dominates the emission in these channels (see
Table~\ref{tab:chan}).
In the 171 channel, the limited areas where $R_A$ significantly
departs from 1 (and is up to 2.9) correspond
to small and narrow features, where the relative contribution of
emission from elements other than Fe is substantial.
These large $R_A$ values are due to ions of elements with high first
ionization potential (FIP), such as \ion{O}{5}/VI, and
\ion{Ne}{4}/V. 

\subsection{193, 211, and 335~\AA{} channels}

For the 193, 211, and 335~\AA{} channels the difference of the synthetic
images assuming photospheric or coronal abundances is twofold.
We find that these channels show large regions where the
$R_A$ varies rather smoothly between $1-1.7$, $1.2-2$ and $1.2-1.5$
for the 193, 211, and 335~\AA{} channels respectively (Figure~\ref{fig:abund} and
Movie~5). This is because these channels contain significant
contributions from non-Fe ions throughout the field of view (as shown
in the above).

In addition, there are also small and narrow features where $R_A$ significantly
departs from 1. These are regions where the relative contribution of
emission from elements other than Fe is very substantial. This is
similar to the very narrow features seen in the 171~\AA{} channel.
In the 193, 211, and 335~\AA{} channels $R_A$ is as high as
2.9, 3.2, and 3.4 respectively. 
These large $R_A$ values are due to ions of elements with high first
ionization potential (FIP), such as \ion{O}{5}/VI, and
\ion{Ne}{4}/V. 
In the case of channels 211 and 335~\AA{}, we observe
significant variability in time, e.g. with the maximum $R_A$ value going down
to 2.9 and 2.3 at time 1200 and 1460~s respectively. This is 
because, at later times in the evolution the atmosphere is
hotter (t=1460~s) and the relative contribution of \ion{O}{0} ions is smaller.

\subsection{131, and 304~\AA{} channels}

For the 131~\AA{} channel, at all instants, the intensity calculated with  
the photospheric abundances is between 2.5 and 3.3 times smaller than 
the intensity calculated with the coronal abundances (see Movie~5). 
We note that $R_A$ does not reach values close to 1 at any location and
time.  The reason is that in this channel the contribution of the high FIP
elements O and Ne to the total emission is important, and the
emission of these non-dominant ions (\ion{O}{6}, and
\ion{Ne}{6}/VII) have characteristic temperatures of formation similar to
the dominant ion, \ion{Fe}{8}. So the temperature dependence of the
relative contribution of the different elements is less pronounced in
the 131~\AA{} channel than in the other channels.

Finally, for the 304~\AA{} channel, at all three instants, the intensity
calculated assuming photospheric abundances ranges from 0.4 to 1 times
the intensity calculated using coronal abundances (see Movie~5). 
Values significantly lower than 1 are due to the contribution of
\ion{Mg}{8} and \ion{Si}{8}/IX/XI to the emission observed in this channel. 

In contrast to the 94, 171, 193, 211, and 335~\AA{} channels, the
131 and 304~\AA{} channels do not show evidence of the narrow features
with high $R_A$ values. This is likely due to two reasons. The dominant ions of
these channels are formed at transition region temperatures, so that
emission from non-dominant ions that have the same formation
temperature as the dominant ion (i.e., transition region) will form in the same narrow,
transition region features as the dominant ions. As a result, $R_A$ maps will
appear relatively flat in the spatial dimension, and lack narrow
spatial features. Or alternatively, if non-dominant ions have higher
formation temperatures, they will tend to form coronal features, which
are typically wider and smoother in the spatial dimension (and thus in
the $R_A$ maps).

\section{Electron density effects}\label{sec:elec}

\begin{figure}
  \includegraphics[width=0.7\textwidth]{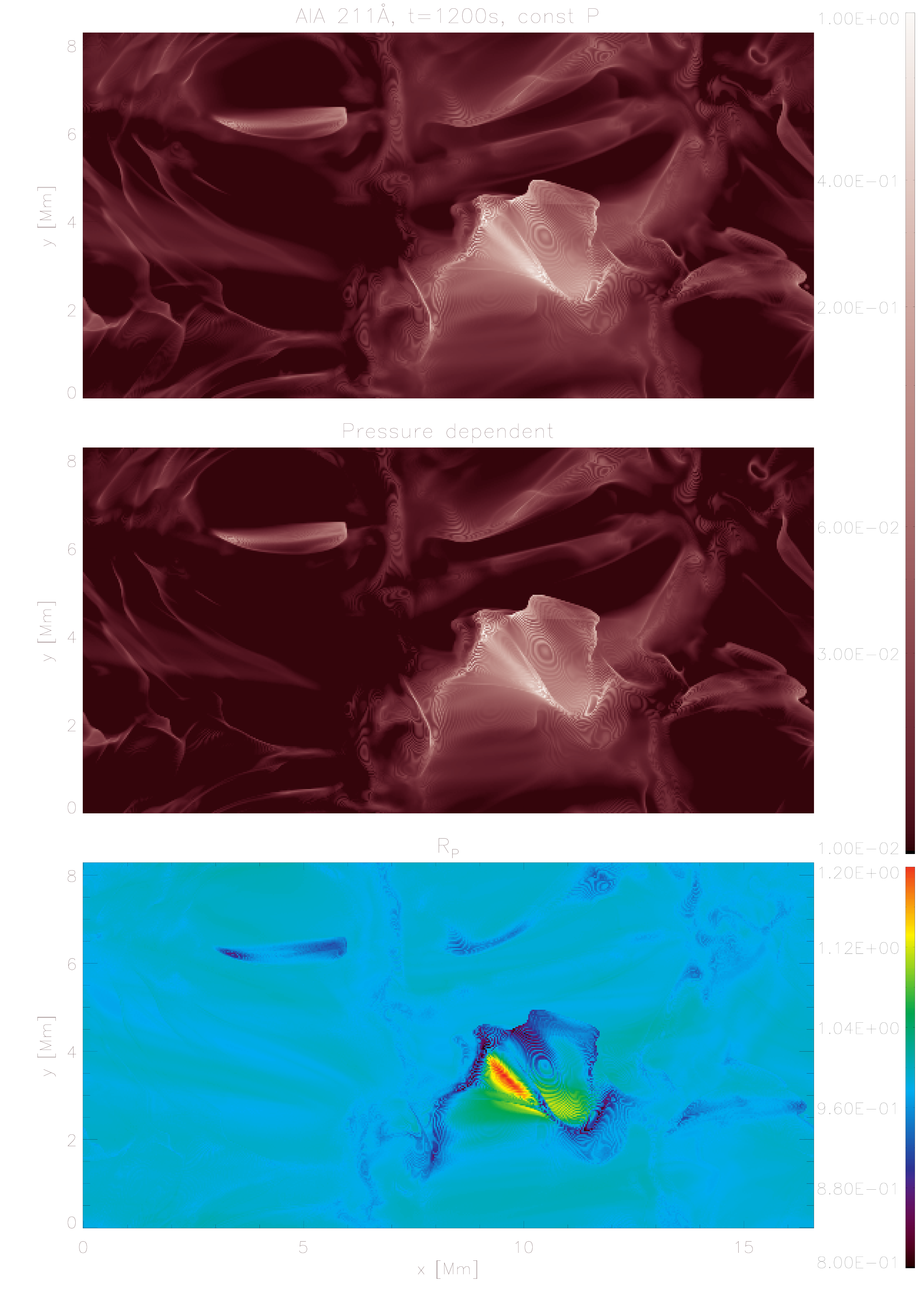}
 \caption{\label{fig:press} A comparison of the synthesized intensity considering 
 constant pressure and taken into account the $G(T,n_{\rm e})$ 
 for the 211~\AA{} channel at time 1200~s are shown in the top and middle panels. 
 The color-scheme is shown in the top-right colorbar in logarithmic scale.
The ratio between the former and the later  ($R_P$) 
is shown in the bottom panel. The color-scheme is shown in the bottom-right colorbar.}
\end{figure}

The line emissivities are generally a function of both temperature and
pressure (or density) of the plasma. Since most lines depend to a much
lesser extent on density than temperature, their emission is often
computed assuming the emissivity $G(T)$, for constant pressure.
Here we investigate the effect of density in the AIA EUV passbands by
comparing the synthetic observations obtained from our 3D dynamic
simulation using $G(T)$ and $G(T,n_{\rm e})$ (see Section~\ref{sec:synthetic}
for a detailed description of the methods to calculated synthetic emission).
In the same manner as the previous section we show here one example
(Figure~\ref{fig:press}) and refer the reader to the online Movie~6. 
Figure~\ref{fig:press} shows the synthesized 211~\AA{} emission using
$G(T)$ (for $P_{\rm e}=10^{15}$~cm$^{-3}$ K; top panel), or
$G(T,n_{\rm e})$ (middle
panel), and their ratio, $R_P$, in the bottom panel.  In the previous
sections, as noted, we have used $G(T,n_{\rm e})$.

\subsection{131, and 335~\AA{} channels}

It is interesting to observe that the ratio $R_P$ does not show 
significant departures from 1 in the 131 and
335~\AA{} channels, indicating that the lines contributing to the
emission in these channels are not density sensitive. For instance, 
the most important lines in channel 335~\AA\ are not density sensitive. 
In contrast, channel 131~\AA\ has two rather important spectral lines 
which are density sensitive: \ion{Fe}{8} 130~\AA{} and \ion{Fe}{8} 
131~\AA{}. However, the dependence with the density is in opposite
directions, and as 
a result the channel is not particularly sensitive to the density.

\subsection{94,  171, 193, 211, and 304~\AA{} channels}

In the 94, 171, 193, and 211~\AA channels, $R_P$ has values
different from 1 in a few structures, especially at later times (t=1200
and 1460~s). Some of these features are common to all these channels,
such as the strong emergent loop in the center of the field of view at
time 1200~s. However, $R_P$ has different values in different regions
for each channel. For instance, in channel 94~\AA{}, $R_P$ is larger
($R_P\approx 1.4$) in the center of the emergent structure. In the
surroundings of this region, $R_P$ decreases slowly until it gets
near to 1. In channel 193~\AA{}, $R_P$ is maximum in the left side
of this emergent structure and below 1 at the boundaries of this
structure. However, other features observed in the $R_P$ maps appear
in some channels but not in others. This can be easily seen in the bottom
part of the image at time 1460~s (in red) in the 211~\AA{} channel),
which is not visible in the other channels.

This difference between the channels indicates that various lines which are density 
sensitive contribute considerably to these channels. 
The 94~\AA{} channel has one spectral line which is rather 
strongly influenced by the density and contributes 
considerably to the emission, i.e., \ion{Fe}{10} 94~\AA{}.
The 171~\AA{} channel has a strong contribution from the spectral 
line \ion{Fe}{9} 171~\AA{} which has some dependence on the density. 
In addition, other lines which contribute less to the emission 
have stronger dependence with the density, i.e., \ion{Fe}{10} 170~\AA{}, 
\ion{O}{5} 171~\AA{}, and \ion{Ni}{14} 171~\AA{}. 
The 193~\AA{} channel has several spectral lines which contribute to the
emission and are density dependent, i.e., several spectral lines 
coming from \ion{Fe}{12} within the range of wavelength 186-191~\AA{}, 
\ion{O}{5} 192~\AA{}, \ion{Fe}{10} 193~\AA{}, \ion{Fe}{11} 188~\AA{} and
several spectral lines coming from \ion{Fe}{9} 
within the range of wavelength 182-189~\AA{}.
The 211~\AA{} channel has some spectral lines which contribute to the
emission and are strongly density dependent, i.e., \ion{O}{5} 215~\AA{}, 
\ion{Fe}{14} 211~\AA{}, \ion{O}{4} 208~\AA{}, \ion{Fe}{13} 202~\AA{},
\ion{Fe}{12} 211~\AA{}, and \ion{Fe}{10} 207~\AA{}.
Finally, the 304~\AA{} channel has faint contributions from a few spectral lines
that are strongly density dependent: \ion{Si}{8} 316~\AA{}, and 
\ion{N}{4} 332~\AA{}. Therefore, since all of these channels
show various spectral lines which have density dependence, the
$R_P$ maps shows values that are different from 1 in different regions.

\begin{deluxetable}{cccc}
\tablecaption{\label{tab:chan} Synthesized SDO/AIA channels}
\tablehead{
\colhead{SDO/AIA channel} &\colhead{Dominant ion(s)} & \colhead{Temperature} & \colhead{Cor/phot abund} }
\startdata
 94~\AA &  \ion{Fe}{10}/XVIII &   $10^6/6\,10^6$~K & 3.981 \\
 131~\AA &  \ion{Fe}{8}/XX &  $6.\,10^5/10^7$~K & 3.981\\
 171~\AA &  \ion{Fe}{9}/X &   $7.\,10^5/10^6$~K & 3.981\\
 193~\AA &  \ion{Fe}{12}/XXIV & $1.2\,10^6/2.\,10^7$~K & 3.981 \\
 211~\AA &  \ion{Fe}{14} &   $2\,10^6$~K & 3.981\\
 304~\AA &  \ion{He}{2} &   $10^5$~K &  0.933\\
 335~\AA &  \ion{Fe}{16} &    $2.5\,10^6$~K & 3.981\\
\enddata
\tablecomments{The synthesized SDO/AIA channels in this work are listed in the first column. The 
dominant ion(s) for each channel, the formation temperature(s) of the dominant ion(s) and the 
ratio between the abundance of the dominant ion(s) assuming the coronal datasets and photospheric datasets are listed 
in the second, third, and fourth columns.}
\end{deluxetable}

\begin{deluxetable}{ccc}
\tablecaption{\label{tab:cross} Selection of $y$-possitions Movie~2}
\tablehead{
\colhead{plot in movie~2} & \colhead{style and color} &\colhead{criteria}   }
\startdata
 1-3a & solid red &  $(I_{NDL}/I_D)I_T$  \\
 1-3b & dashed red &   $I_{NDL}/I_D$  \\
1-3c & solid green & $(I_{NDH}/I_D)I_T$ \\
1-3d & dashed green &  $I_{NDH}/I_D$  \\
\enddata
\tablecomments{Each panel (labels listed in the first column of 
this table) in Movie~2 shows results for a different
 $y$-position in the field of view. These locations are marked with crosshairs in the panels 
1-3D in Movie~1, the color and style of each crosshair is given in
the second column of this table. The chosen locations are such that the
criterium listed in the third column of this table is maximized. }
\end{deluxetable}

\end{document}